\documentclass[10pt]{article}

\usepackage[
    letterpaper,
    margin=1in
]{geometry}

\usepackage[T1]{fontenc}
\usepackage{newtxtext}
\usepackage{microtype}
\usepackage{booktabs}
\usepackage{multirow}
\usepackage{amsmath}
\usepackage{amssymb}
\usepackage{bm}

\usepackage{booktabs}
\usepackage{multirow}
\usepackage{tabularx}
\usepackage{longtable}
\usepackage{threeparttable}
\usepackage{siunitx}
\usepackage{ragged2e}
\newcolumntype{L}[1]{>{\RaggedRight\arraybackslash}p{#1}}
\newcolumntype{Y}{>{\RaggedRight\arraybackslash}X}

\usepackage{graphicx}
\usepackage{float}
\usepackage{placeins}
\usepackage{subcaption}
\usepackage{xcolor}
  \definecolor{HumeInk}{HTML}{353535} % brand dark — text/spines/rules
  \definecolor{HumeBlue}{HTML}{6E9EE8}% Hume SV blue (used for role: speaker-verification / secondary)
  \definecolor{HumeLightBlue}{HTML}{EEF3FB}% ~90 % white tint of HumeBlue
  \definecolor{HumeRule}{HTML}{E6E6E1}% Hume warm gray — rules, gridlines
  \definecolor{HumeGray}{HTML}{6B6B6B}% Hume mid gray — captions, error bars, secondary text
\usepackage{enumitem}

\usepackage[numbers,sort&compress]{natbib}
\usepackage[
    colorlinks=true,
    linkcolor=HumeBlue,
    citecolor=HumeBlue,
    urlcolor=HumeBlue
]{hyperref}
\usepackage[nameinlink,capitalise,noabbrev]{cleveref}

\usepackage{titlesec}
\usepackage{titling}
\usepackage[font=small,labelfont=bf,labelsep=period]{caption}
\usepackage[most]{tcolorbox}
\titleformat{\section}{\Large\bfseries\color{HumeInk}}{\thesection}{0.65em}{}
\titleformat{\subsection}{\large\bfseries\color{HumeInk}}{\thesubsection}{0.65em}{}
\titleformat{\subsubsection}{\normalsize\bfseries\color{HumeInk}}{\thesubsubsection}{0.65em}{}
\titlespacing*{\section}{0pt}{2.0ex plus 0.6ex minus 0.2ex}{0.8ex}
\titlespacing*{\subsection}{0pt}{1.5ex plus 0.4ex minus 0.2ex}{0.5ex}
\titlespacing*{\subsubsection}{0pt}{1.0ex plus 0.3ex minus 0.2ex}{0.35ex}
\newcommand{\rwvoiceeq}{\textsc{RW-Voice-EQ}}
\newcommand{\humelogo}{%
  \IfFileExists{figures/Hume_Logo_Black.png}{%
    \includegraphics[width=1.5in]{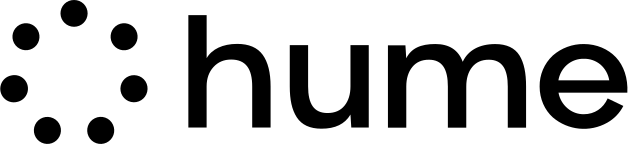}%
  }{\textbf{Hume AI}}%
}
\pretitle{\noindent\humelogo\par\vspace{1.8em}\begin{center}\LARGE\bfseries\color{HumeInk}}
\posttitle{\par\vspace{0.55em}{\color{HumeRule}\rule{0.68\linewidth}{0.6pt}}\end{center}}
\preauthor{\begin{center}\normalsize\color{HumeGray}\begin{minipage}{0.92\linewidth}\centering}
\postauthor{\end{minipage}\end{center}\vspace{-0.6em}}
\predate{}
\postdate{}
\renewenvironment{abstract}{%
  \begin{center}
  \begin{tcolorbox}[enhanced,width=0.94\linewidth,colback=HumeLightBlue,colframe=HumeRule,boxrule=0.5pt,arc=1.5pt,left=9pt,right=9pt,top=7pt,bottom=7pt]
  \small\noindent\textbf{Abstract.}\space
}{%
  \end{tcolorbox}
  \end{center}
}

\AtBeginDocument{\color{HumeInk}}
\renewcommand{\arraystretch}{1.12}
\setlist[itemize]{leftmargin=2.5em, itemsep=0.25em, topsep=0.25em}

\title{\rwvoiceeq{} \textsc{Bench}: A Real World Benchmark for \\ Evaluating Voice AI Systems}

\author{
David Ayllon\textsuperscript{*}, 
Alice Baird\textsuperscript{*}, 
Jeffrey Brooks\textsuperscript{*}, 
Franc Camps-Febrer\textsuperscript{*}, 
Jakub Piotr C\l{}apa\textsuperscript{*}, \\
Theo Lebryk\textsuperscript{*}, 
Jens Madsen\textsuperscript{*}, 
Olya Ossipova\textsuperscript{*}, 
Sharath Rao\textsuperscript{*}, 
Hoon Shin\textsuperscript{*}, 
Tigran Soghbatyan\textsuperscript{*}, \\
Georg Streich\textsuperscript{*}, 
Rashish Tandon\textsuperscript{*}, 
Panagiotis Tzirakis\textsuperscript{*}
\\[0.45em]Hume AI Research
}

\date{}

\begin{document}

\maketitle
\begingroup

\renewcommand{\thefootnote}{*}

\footnotemark
\footnotetext{All authors contributed equally, share first authorship, and are listed in alphabetical order.

Corresponding authors: \texttt{panagiotis@hume.ai}, \texttt{alice@hume.ai}.}

\endgroup
\begin{abstract}
Current voice AI benchmarks typically evaluate isolated capabilities such as speech intelligibility, word error rate, or text-based dialogue quality, but they rarely test whether systems harness the acoustic information that distinguishes spoken language from its textual representation. To this end, we introduce the Real World Voice EQ Bench~\footnote{\href{https://huggingface.co/spaces/HumeAI/rw-voice-eq}{\texttt{huggingface.co/spaces/HumeAI/rw-voice-eq}}}, a multidimensional benchmark for evaluating voice AI across text-to-speech (TTS), speech-to-speech (STS), speech understanding (SU), and automatic speech recognition (ASR). Our evaluations indicate that performance is highly dimension-specific. For TTS, naturalness, expressiveness, identity stability, and reliability are largely independent evaluation dimensions. For STS, access to audio does not guarantee use of vocal affect, and some agents remain largely transcript-driven. For SU, models perform unevenly across paralinguistic tasks. For ASR, real world accent, emotion, noise, and conversational conditions expose failures that are not captured by established clean-speech benchmarks. Together, these results show that voice AI should be evaluated as a profile of acoustic, expressive, interactional, and robustness capabilities rather than by a single aggregate score.
\end{abstract}

% \tableofcontents

\section{Introduction}

Speech is one of the primary channels of human communication. It carries linguistic content, but also a rich set of acoustic cues that shape how that content is understood \cite{couperkuhlen2018interactional,barthweingarten2010prosody,gussenhoven2021handbook,laver1980voicequality}. When people speak to one another, they do not exchange words alone: they signal confidence or uncertainty, express affect, adapt to urgency, recover from noise and interruption, and infer social and emotional context from how speech is produced \cite{scherer1979socialmarkers,sacks1974turntaking}. A sentence such as ``I'm fine'' can function as reassurance, irritation, resignation, sarcasm, or distress depending on its prosody---including timing, loudness, stress, and intonation---as well as vocal effort and voice quality. These distinctions are often absent from the transcript, yet they are central to how human listeners interpret intent, stance, emotion, and the appropriate next action.

Voice AI systems are increasingly moving from narrow transcription and synthesis interfaces toward full spoken interaction \cite{chen2026voicebench,liu2025vocalbench}. Modern systems must not only recognize or generate the correct words, but also produce expressive and speaker-consistent speech, respond appropriately to vocal tone, extract paralinguistic information, and remain effective under conversational and acoustic stress \cite{shah2024speechrobustbench,liu2026speechparaling}. Evaluating these systems therefore requires measuring whether they perceive, generate, and act on information carried by the spoken signal rather than assessing text-level competence alone.

Evaluation has expanded substantially within individual areas of voice AI. Text-to-speech (TTS) evaluation has progressed from standardized subjective-quality protocols \cite{itu1996p800,itu2015bs1534} and shared synthesis evaluations such as the Blizzard Challenge \cite{black2005blizzard}, through automatic mean-opinion-score prediction challenges such as VoiceMOS \cite{huang2022voicemos,huang2024voicemos}, to benchmarks such as InstructTTSEval, EmergentTTS-Eval, and SpeechParaling-Bench, which probe instruction following, role-play, expressiveness, prosody, and pronunciation \cite{huang2025instructttseval,manku2025emergenttts,liu2026speechparaling}. Earlier spoken-dialogue evaluation frameworks such as PARADISE separated task success from interaction costs \cite{walker1997paradise}. More recent spoken-input and spoken-interaction benchmarks---including SD-Eval, VoiceBench, URO-Bench, S2S-Arena, and VocalBench---evaluate combinations of spoken understanding, reasoning, instruction following, paralinguistic awareness, conversational ability, and acoustic response quality \cite{ao2024sdeval,chen2026voicebench,yan2025urobench,jiang2026s2sarena,liu2025vocalbench}. AIR-Bench, AudioBench, and AHELM provide broader evaluations of audio-language understanding \cite{yang2024airbench,wang2025audiobench,lee2025ahelm}. For automatic speech recognition (ASR), Common Voice and FLEURS support multilingual and speaker-diverse evaluation, while the CHiME challenges and Speech Robust Bench target noisy, conversational, and corrupted speech \cite{ardila2020commonvoice,conneau2023fleurs,watanabe2020chime6,shah2024speechrobustbench}.

Despite this progress, evaluation remains fragmented across system interfaces, capability definitions, and scoring protocols. Existing benchmarks rarely determine whether performance improves specifically because a model has access to acoustic information, whether semantic response quality diverges from the quality of the generated voice, or whether aggregate scores conceal dimension-specific failures. Consequently, a system may sound natural while failing to perform an expressive role, generate an appropriate response from a transcript while ignoring vocal affect, or achieve strong average ASR performance while failing on particular accents, emotional speech, background speakers, or noisy conversational conditions \cite{manku2025emergenttts,ao2024sdeval,jiang2026s2sarena,shah2024speechrobustbench}.

To address this gap, we introduce the Real World Voice-EQ Bench (\rwvoiceeq{}), a multidimensional benchmark for evaluating voice AI systems across four domains: text-to-speech (TTS) generation, speech-to-speech (STS) interaction, speech understanding (SU), and automatic speech recognition (ASR). \rwvoiceeq{} is organized around a central distinction between linguistic content and paralinguistic acoustic cues. Linguistic content captures what is said and can generally be approximated from a transcript. Paralinguistic cues capture how something is said and who is speaking, including speaker states and traits, prosody, timbre, speaker identity, vocal effort, affect, hesitation, and audible context. The benchmark therefore tests whether voice systems rely only on lexical content or use acoustic information that transcripts do not preserve.

\rwvoiceeq{} evaluates TTS systems on acting and role fit, expressiveness, speaker identity, language stability, reliability, long-form speaker stability, and acoustic quality. STS systems are evaluated on emotion understanding, emotion alignment, expressivity robustness, voice naturalness, and problem redirection. Speech-understanding models are evaluated on speech-emotion recognition, speaker verification, and synthetic-speech detection, while ASR systems are evaluated on transcription robustness under accented, emotional, noisy, and conversational conditions. Together, these evaluations characterize real world voice-system performance and test whether models use the acoustic, expressive, and interactional cues that shape spoken communication.

The contributions of this paper are:

\begin{itemize}
    \item We introduce Real World Voice-EQ Bench (\rwvoiceeq{}), a multidimensional benchmark for evaluating voice AI systems across text-to-speech generation, speech-to-speech interaction, speech understanding, and automatic-speech-recognition robustness.
    \item We define a multidimensional evaluation taxonomy that separates linguistic-content performance from paralinguistic and acoustic-interactional competence, covering naturalness, expressiveness, speaker identity, reliability, audio sensitivity, speaker verification, synthetic-speech detection, and transcription robustness.
    \item We provide targeted TTS evaluations for real world production-relevant generation capabilities, including expressive and role-appropriate speech, voice identity preservation, language stability, long-form stability, content reliability.
    \item We introduce expression-driven STS evaluations that test whether voice agents utilize the acoustic and paralinguistic information present during interaction with humans, including vocal affect, tone, hesitation, urgency, hostility, degraded audio, problem redirection, as well as evaluating the naturalness of the agent’s own spoken response under stress.
    \item We evaluate speech-understanding models for emotion recognition, speaker verification, and synthetic-speech detection.
    \item We analyze ASR robustness under accented, emotional, noisy, and conversational speech, showing that clean-speech transcription benchmarks miss production-relevant failure modes.
    \item We analyze agreement between our pool of human raters and speech-language model judges to understand places in which automatic judges are reliable and where they continue to fall short. 
    \item We show that voice AI  performance is highly dimension-specific: systems that are strong on naturalness, expression, speaker identity, reliability, or speaker verification are not necessarily strong in other domains. This supports reporting voice AI performance as a dimension profile rather than a single aggregate score.
\end{itemize}

\section{\rwvoiceeq{} Benchmark}

\begin{figure}[H]
    \centering
    \includegraphics[width=0.7\linewidth]{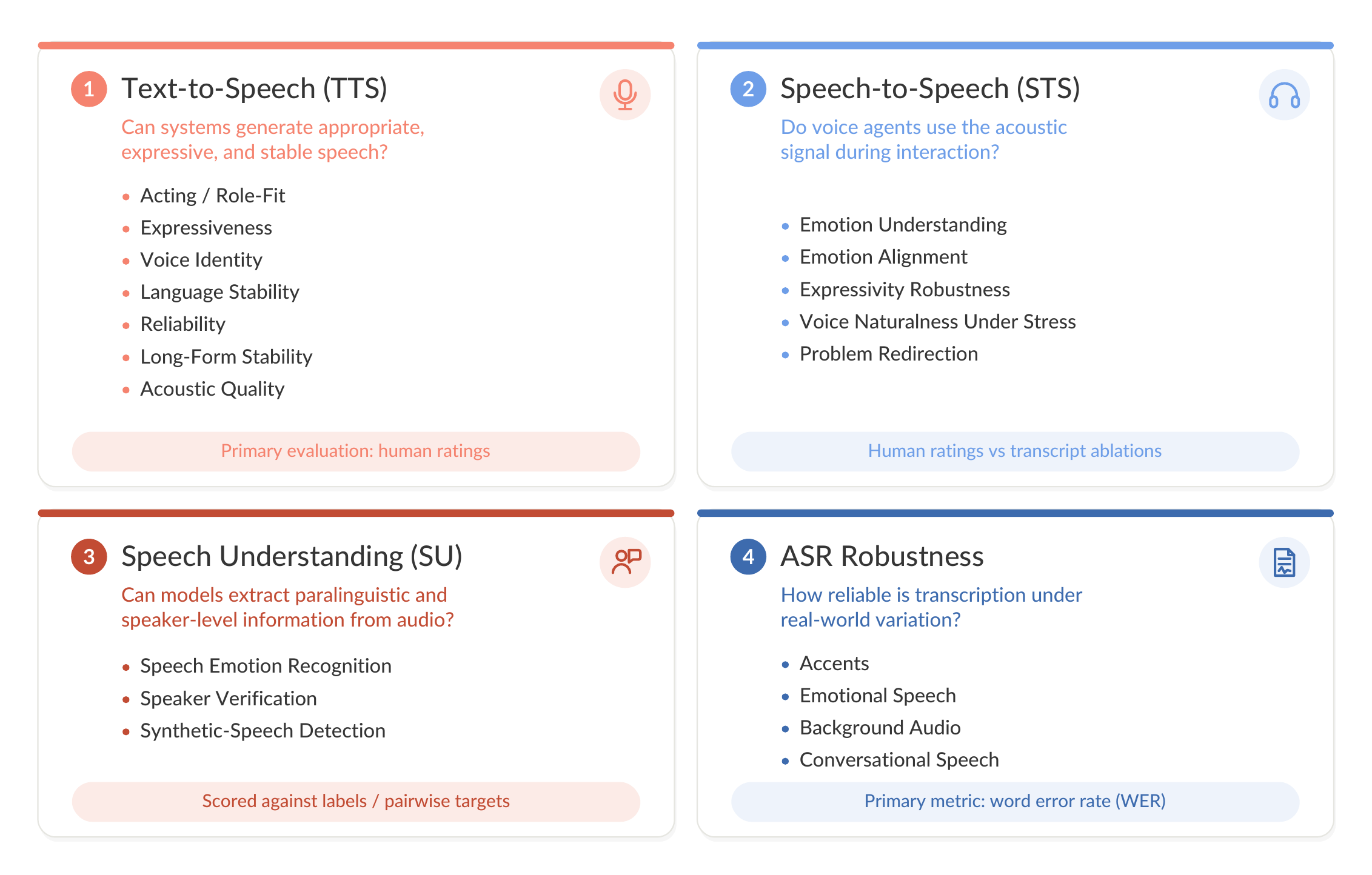}
 \caption{Overview of the \rwvoiceeq{} benchmark. \rwvoiceeq{} evaluates
  voice AI along four domains: \textbf{(1) Text-to-Speech (TTS)},
  \textbf{(2) Speech-to-Speech (STS)}, \textbf{(3) Speech Understanding (SU)},
  \textbf{(4) ASR Robustness}. Together the four domains characterize a
  voice system as a capability profile rather than a single aggregate score.}
    \label{fig:benchmark-overview}
\end{figure}

\subsection{Benchmark Design}

\rwvoiceeq{} is designed to evaluate whether voice AI systems remain effective under realistic conditions. Rather than treating speech as a single capability, the benchmark separates four domains that arise in real world voice AI systems (\cref{fig:benchmark-overview}): generating speech, interacting through speech, understanding paralinguistic and speaker-level information, and recognizing spoken content under acoustic variation.

These domains correspond to text-to-speech generation, speech-to-speech interaction, speech understanding, and automatic speech recognition robustness. Text-to-speech (TTS) evaluates systems that generate speech from text, style instructions, or voice-design inputs. Speech-to-speech (STS) evaluates interactive voice agents that receive spoken input and produce spoken responses. Speech understanding (SU) evaluates models that receive speech or audio clips as input and return labels, scores, or text descriptions of the audio. Automatic speech recognition (ASR) evaluates transcription systems under acoustic and sociolinguistic variation.

The taxonomy is dimension-oriented rather than model-oriented. A system is evaluated within a domain only when it supports the required input-output interface and task protocol. Each evaluation targets a specific dimension, uses a fixed item set, and defines its rating scale or scoring rule before outputs are reviewed. Results are reported separately by construct and are not pooled across constructs, because different tasks measure different capabilities and often use different scales. This design is intentional: \rwvoiceeq{} is intended to produce diagnostic profiles of model behavior rather than a single global voice-quality score.

\textbf{Text-to-Speech (TTS) domain}. A TTS system must do more than sound natural in isolated English sentences. It must preserve speaker identity, follow expressive and pragmatic instructions, read task-critical content correctly, switch between languages without losing intelligibility or speaker consistency, remain natural across varied text, and sustain appropriate delivery in genre-specific performances. For these reasons, the TTS benchmark contains seven evaluation dimensions:

\begin{itemize}
    \item \textbf{Acting / Role-Fit:} Tests whether the model can adopt a genre-appropriate performative role, such as narrator, public speaker, commentator, comedian, or casual speaker.
    \item \textbf{Expressiveness:} Tests whether the model renders intended affect, pragmatics, prosodic emphasis, sarcasm, emotional transitions, and production-ready expressive tone.
    \item \textbf{Voice Identity:} Tests whether the generated voice remains recognizably the same speaker under emotional, prosodic, physical, register, and transition stressors.
    \item \textbf{Language Stability:} Tests whether the system produces stable, contextually correct pronunciations when linguistic context changes. 
    \item \textbf{Reliability}: Tests whether the model accurately reads task-critical content such as alphanumeric strings, structured data, and pharmaceutical names.
    \item \textbf{Long-Form Stability:} Tests whether the model remains natural, human-like, and speaker-consistent over medium-form and document-scale generation.
    \item \textbf{Acoustic Quality:} Tests whether the model produces clean, stable audio without distracting artifacts, distortions, dropouts, unnatural noise, or quality degradation across varied text conditions.
\end{itemize}

\textbf{Speech-to-Speech (STS) domain.} The STS benchmark evaluates whether voice agents can use spoken input to produce appropriate spoken responses. In real conversations, the transcript is often insufficient: a user may sound hesitant while saying yes, panicked while asking a simple question, hostile while using polite words, or unclear because the channel is degraded. A capable voice agent should not treat speech as text with audio attached. It should use vocal tone, urgency, hesitation, affect, and acoustic quality when deciding how to respond. Accordingly, the STS benchmark evaluates spoken dialogue systems in situations where the input audio itself matters. It covers five task designs:
\begin{itemize}
    \item \textbf{Emotion Understanding:} Tests whether spoken affect improves performance beyond a transcript-only anchor using audio-conditioned and transcript-only ablations.
    \item \textbf{Emotion Alignment:} Tests whether the agent responds appropriately when the user’s words appear to confirm an action but their vocal tone signals hesitation, fear, anger, confusion, or reluctance.
    \item \textbf{Expressivity Robustness:} Tests whether the agent remains useful and behaviorally stable under degraded audio, hostile users, and urgent users.
    \item \textbf{Voice Naturalness:} Tests whether the agent’s own voice remains natural under the same difficult conversational conditions.
    \item \textbf{Problem Redirection:} Tests whether the agent maintains composure with a hostile user and redirects the conversation back to the underlying problem.
\end{itemize}

\textbf{Speech Understanding (SU) domain.} The SU benchmark evaluates whether models can extract perceptual and speaker-level information from audio clips. These tasks are not ASR or transcription evaluations. Instead, they target audio-understanding capabilities that are relevant both to voice-agent behavior and to benchmark validity: recognizing emotion, comparing affective intensity, matching speakers, and distinguishing real from synthetic speech. Accordingly, the SU benchmark evaluates whether models can perceive properties of speech across four task types:
\begin{itemize}
    \item \textbf{Speech Emotion Recognition:} Tests whether the model can recognize dominant emotion and intensity of a clip
    \item \textbf{Speaker Identification:} Tests whether the model can decide if two clips are from the same speaker.
    \item \textbf{Synthetic-speech Detection:} Tests whether the model can distinguish human speech from generated speech.
\end{itemize}

\textbf{ASR Robustness domain.} The ASR benchmark evaluates whether transcription remains reliable under realistic acoustic and sociolinguistic variation. Recognition accuracy on clean or semi-controlled speech does not necessarily imply robustness to accented, emotional, noisy, or conversational speech. The ASR evaluations therefore test whether systems preserve lexical reliability when speech conditions differ from standard benchmark settings.
\begin{itemize}
    \item \textbf{Emotions:} Tests how well the ASR model performs under various expressive states, specifically grouped across the arousal / valence circumplex -  positive valence, negative low arousal and negative high arousal.
    \item \textbf{Accents:} Tests how well the ASR model generalizes across speakers with diverse English accents, grouped under three fairness-oriented taxonomies: Kachru’s Three Circles (Native / Second-language / Foreign), Native vs. Non-native, and Standard American vs. Non-standard American.
    \item \textbf{Background Audio:} Tests how well the ASR model performs in the presence of competing background sounds, specifically grouped into music (predominantly instrumental background audio) and noise (e.g., crowd chatter, traffic, restaurant ambience), while speech remains in the foreground.
    \item \textbf{Conversational:} Tests how well the ASR model transcribes speech in realistic conversational environments with multiple speakers, including overlapping speech and competing background talkers.
\end{itemize}

\subsection{Evaluation Setup and Methodology}

\begin{figure}[h]
    \centering
    \includegraphics[width=0.7\linewidth]{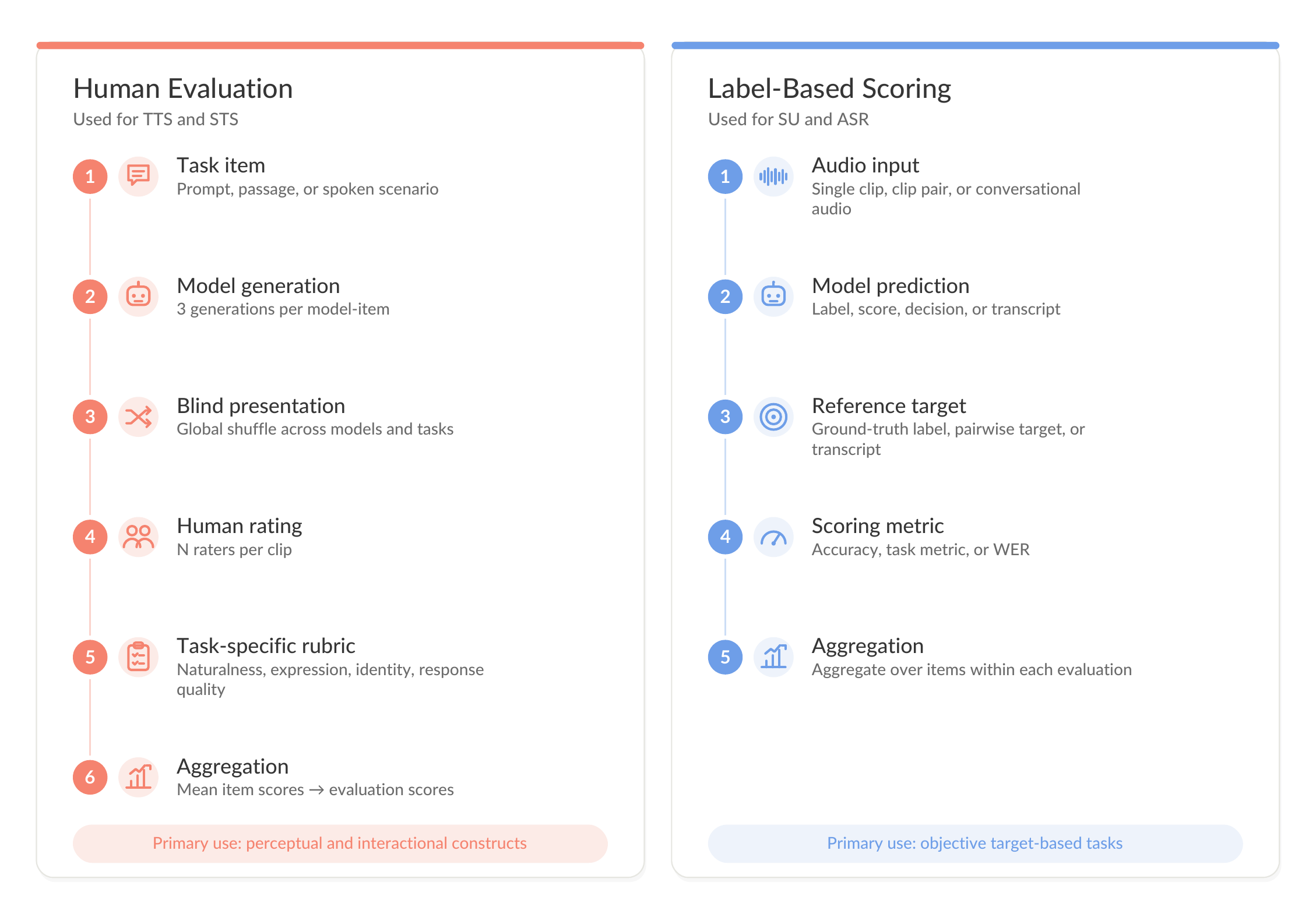}
   \caption{The \rwvoiceeq{} evaluation methodology. \textbf{Human evaluation}
  (TTS, STS) generates audio per model-item, presents it to multiple
  raters against a task-specific rubric, and aggregates into per-evaluation
  scores. \textbf{Label-based scoring} (SU, ASR) compares model predictions
  against ground-truth labels, pairwise targets, or reference transcripts
  using accuracy or word error rate. The two paradigms respectively cover
  perceptual/interactional constructs and objective target-based tasks.}
    \label{fig:evaluation-methodology}
\end{figure}

\rwvoiceeq{} follows a common design principle across all benchmark families: models are evaluated on the task they are meant to perform, rather than on an indirect proxy. For TTS and STS, the target output is audio intended to be perceived by a listener, so quality is judged directly by human raters using task-specific rubrics. For SU and ASR, the target output is a decision or transcript with a predefined target, so model predictions are scored against labels, pairwise targets, or reference transcriptions.

This yields two main evaluation paradigms. Human evaluation is used when the relevant outcome is inherently perceptual, such as naturalness, expressiveness, speaker identity, emotional appropriateness, or conversational response quality. Label-based scoring is used when the task has a well-defined target, such as identifying an emotion, comparing emotional intensity, matching speakers, detecting synthetic speech, or transcribing spoken content. In both paradigms, each evaluation uses a fixed item set, a predefined rating scale or label space, and a scoring rule specified before model outputs are reviewed.

\textbf{Human Studies for TTS and STS.} Human raters were recruited through Hume's Study Runner API platform and filtered to primary English speakers residing in the United States, Great Britain, or Canada. Detailed demographic statistics are provided in~\cref{tab:participant-demographics}. All raters provided informed consent and were financially compensated for their participation. 

Across the development of \rwvoiceeq{}, we collected over one million individual human ratings. The final benchmark comprises 785,679 ratings for TTS evaluations and 48,053 ratings for STS evaluations; the remaining ratings were collected during iterative benchmark refinement, including studies used to assess discriminative validity, remove redundant evaluations, and improve real-world representativeness.

Language-switching evaluations apply a matching native-speaker filter for each target language. Each clip is rated by three independent raters. A session presents up to ten clips, drawn as a globally shuffled uniform random subset across models and scenarios, so no rater sees a single-model or single-category block.

Ratings use a five-point Likert scale unless otherwise specified. The question wording is specific to each evaluation: for example, voice-identity evaluations ask whether two clips sound like the same speaker, while expression evaluations ask whether the voice matches an intended emotional target. Raters are instructed to use headphones and judge only what they hear. Prompt text or scenario context is shown when needed, but the audio is the basis for the rating. Rater assignment is blind to model identity.

\textbf{Label-Based Scoring for SU and ASR.} For SU, the model receives one or more audio clips and outputs a label, score, or decision. Ground truth is drawn from corpus annotations or predefined pairwise targets, including emotion labels, same-speaker pairs, and real/synthetic labels. For ASR robustness, model transcriptions are compared against reference transcripts using word error rate or the relevant task-specific recognition metric. No human rater is involved in these scoring loops; the model response is compared directly against the reference target, which has been previously human generated.

\textbf{Eval Dimensions and Factor Scoring.} For TTS and STS individual evals are grouped into latent factors (Evaluation Dimensions).  Each eval has a fixed set of primary rating questions (e.g., Audiobook Narration, Code-Switching - German  and Heteronym Pronunciation) whose responses are averaged into a per-(provider, item) composite. Composites are z-scored within eval and passed through a mixed-effects model, yielding a rater-controlled provider mean per eval that removes systematic rater biases and per-item variance. Reverse-coded questions (artifacts presence) are flipped so that higher is uniformly better. A factor score is the equal-weighted mean of its constituent evals' rater-controlled provider means. To validate the grouping empirically, each eval's provider vector is Spearman-correlated against every factor leaderboard (eval dimension); an eval is treated as fitting the factor with which it correlates most strongly, and any eval whose best-fit correlation falls below 0.30 is flagged as an orphan candidate for regrouping. These eval groupings into factors are verified manually to make sure spurious correlations are ignored and keep the factors logically consistent. 

\section{Preliminary Studies}

Before presenting our main analyses, we first establish two empirical observations that motivate the remainder of the paper. Specifically, we ask two questions: (1) can speech-language models reliably substitute for human evaluators? and (2) where do current open-source benchmarks fall short? We begin by evaluating the extent to which speech-language models agree with human listeners on TTS evaluation, validating their use as scalable judges. We then introduce a set of diagnostic evaluations showing that benchmark optimization is already present in state-of-the-art ASR systems. Together, these preliminary studies establish both the measurement methodology and the motivation for the mechanistic analyses that follow.

\subsection{Human-SLM Agreement}

Speech-language models (SLMs) offer a scalable approach to TTS evaluation, but their reliability as proxies for human judgments remains underexplored. When we compare Human-SLM judgements, we focus on our TTS domain~\footnote{The full SLM judge leaderboard is publicly available at \href{https://huggingface.co/spaces/HumeAI/slm-judge-leaderboard}{\texttt{huggingface.co/spaces/HumeAI/slm-judge-leaderboard}}.}, mainly as human evaluation protocols are substantially more standardized than for STS, providing a cleaner setting for validating SLMs as proxies for human listeners, however future research would include evaluation of SLM judges in STS paradigms.

\begin{figure}[H]
    \centering
    \includegraphics[width=0.5\linewidth]{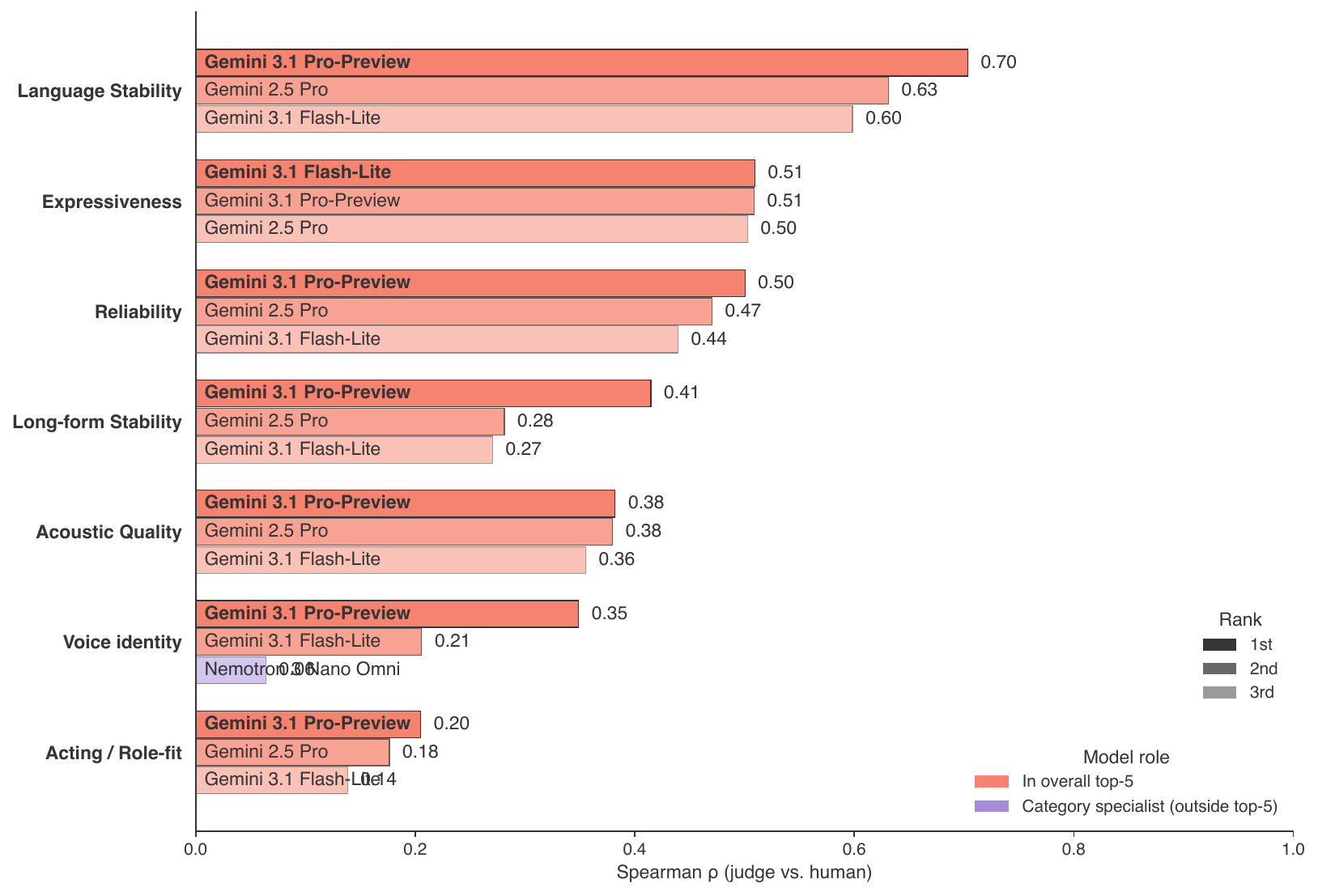}
  \caption{Human--SLM agreement across TTS dimension groups. For each SLM judge and dimension group, agreement is reported as the mean Spearman rank correlation ($\rho$) between SLM judge scores and human ratings across the constituent evaluations. The three highest-scoring judges within each dimension group are shown. Higher values indicate closer agreement with human system rankings.}
    \label{fig:leaderboard-top3-per-category}
\end{figure}

\cref{fig:leaderboard-top3-per-category} shows the mean Spearman’s correlation between human and SLM ratings across TTS evaluation dimensions, averaged across seven publicly available models (left) and the top 3 models across all categories (right). These modes include proprietary frontier models and open-weight models: Gemini 3.1 Pro (preview), Gemini 2.5 Pro, Gemini 3.1 Flash Lite, Gemini 2.5 Flash, GPT Audio 1.5, Kimi Audio-7b Instruct, Nemotron 3 Nano Omni.

The results in \cref{fig:leaderboard-top3-per-category} indicate that agreement is higher for language- and reliability-driven categories, tasks that are graded against a known, statable answer (e.g., was this word pronounced correctly given the context, does the generation match the target emotion, was the correct number read back). When we look closer at the expressive category, correlation is not uniformly high: it depends on how easily that answer can be verified once known. Cases that pair the audio with an explicit target and ask the judge to confirm it, an unambiguous text-inferable emotion, an explicit style instruction, reach some of the highest correlations.

This also reflects the quality of the generations themselves; when looking at the standard deviation between human raters, we see that more ambiguous samples have much lower correlations and this warrants further exploration in future work. In these cases the generations themselves tend to clearly express the target emotion or clearly fail to, so there is little ambiguity in what either a human or a judge should conclude, and the two converge easily as a result.

The size of this effect becomes clear once the evals that supply an explicit textual clue are removed from the category altogether. Averaged across judges, the expression category's mean correlation with human ratings is r = 0.47 when it includes an eval built on unambiguous, text-inferable or explicitly instructed cases; restricting the same category to evals that withhold or complicate that clue cuts the mean roughly in half, to r = 0.25. The apparent reliability of "expression" as a category is therefore driven disproportionately by the presence of a small number of easy, cue-rich evals, rather than to any inherent advantages in the judge's ability to assess expressive constructs.

Performance drops off similarly for constructs with no stated answer at all, including acoustic quality, acting role fit, and voice identity. For voice identity in particular, raters are asked to judge whether the generated voice remains the same speaker when a stressor is applied (e.g. talk very slowly, with  a furious delivery), and on this unaided judgment SLM judges track human ratings especially poorly. We also acknowledge that task-specialist models may outperform general-purpose SLM judges on narrowly defined evaluation tasks where the target construct is well specified (e.g., speaker verification, and speech emotion recognition). A systematic comparison between specialist models, general-purpose SLM judges, and human raters across the evaluation dimensions explored in \rwvoiceeq{} remains an important direction for future work.

This pattern suggests SLM judges are not simply construct-dependent but verification-dependent, i.e., they perform well when a target is given and either the deviation from it is limited or the generation itself is unambiguous, and degrade as soon as a judgment requires genuine, unaided perceptual discrimination on borderline cases, even inside categories, like expression, that look reliable on average. They are useful for screening and for constrained, answer-graded correctness tasks, but human ratings remain necessary for perceptual, open-ended, socially grounded judgments, acting role fit, voice identity, and the harder end of expressive delivery, where no stated target can substitute for listening. \rwvoiceeq{} therefore treats SLM judges as auxiliary evaluators and uses human ratings as the primary signal particularly when the construct depends on listener perception rather than verification against a known target.

\subsection{Evidence of Benchmark Optimization}

Standard ASR leaderboards rely on open-source speech datasets, such as LibriSpeech~\cite{panayotov2015librispeech} and VoxPopuli~\cite{wang2021voxpopuli}. The prevalence of these datasets across benchmarks creates systematic pressure toward optimization that improves reported scores without improving generalizable acoustic capability. We term this benchmark optimization, also colloquially referred to as ``benchmaxxing'': improving benchmark performance by overfitting to arbitrary features of the evaluation dataset rather than by learning to recognize speech.

We designed four complementary tests, each targeting a different area through which a model might score well on a benchmark without genuine acoustic generalization. The tests do not implicate the same models, which is itself the central result: each catches a different optimization strategy.
\begin{itemize}
    \item \textbf{Consensus disagreement.} VoxPopuli contains reference transcripts that are demonstrably incorrect. We identify these errors with a consensus panel of models and confirm them through human verification. Models optimized for the benchmark often reproduce the incorrect reference even when the audio says otherwise.
    \item \textbf{Audio masking.} We silence specific words in the audio, names, proper nouns, and numbers that should be impossible to guess from acoustic context, and test whether models still output the masked word. Genuine transcriptions should not include the masked audio.
    \item \textbf{Orthographic switch.} Many words have multiple phonetically identical spellings (e.g., Mr. vs. mister, color vs. colour). Because audio cannot determine the correct form, matching the benchmark’s specific convention demonstrates that the model can identify which benchmark an audio came from and implicitly applies that benchmark’s convention during transcription. Above-chance matching indicates memorization of reference-text patterns rather than transcription.
    \item \textbf{Synthetic voices.} For segments where benchmark-optimized models reproduce an incorrect reference, we synthesize the correct transcript with two voices: a generic voice and one matched to the original speaker. If a model transcribes the generic voice correctly but repeats the incorrect reference for the matched voice, it suggests audio-level memorization by associating specific voices or benchmark audio signatures with a specific transcript.
\end{itemize}

\begin{figure}[!t]
    \centering
    \includegraphics[width=0.5\linewidth,trim=0 0 0 90pt,clip]{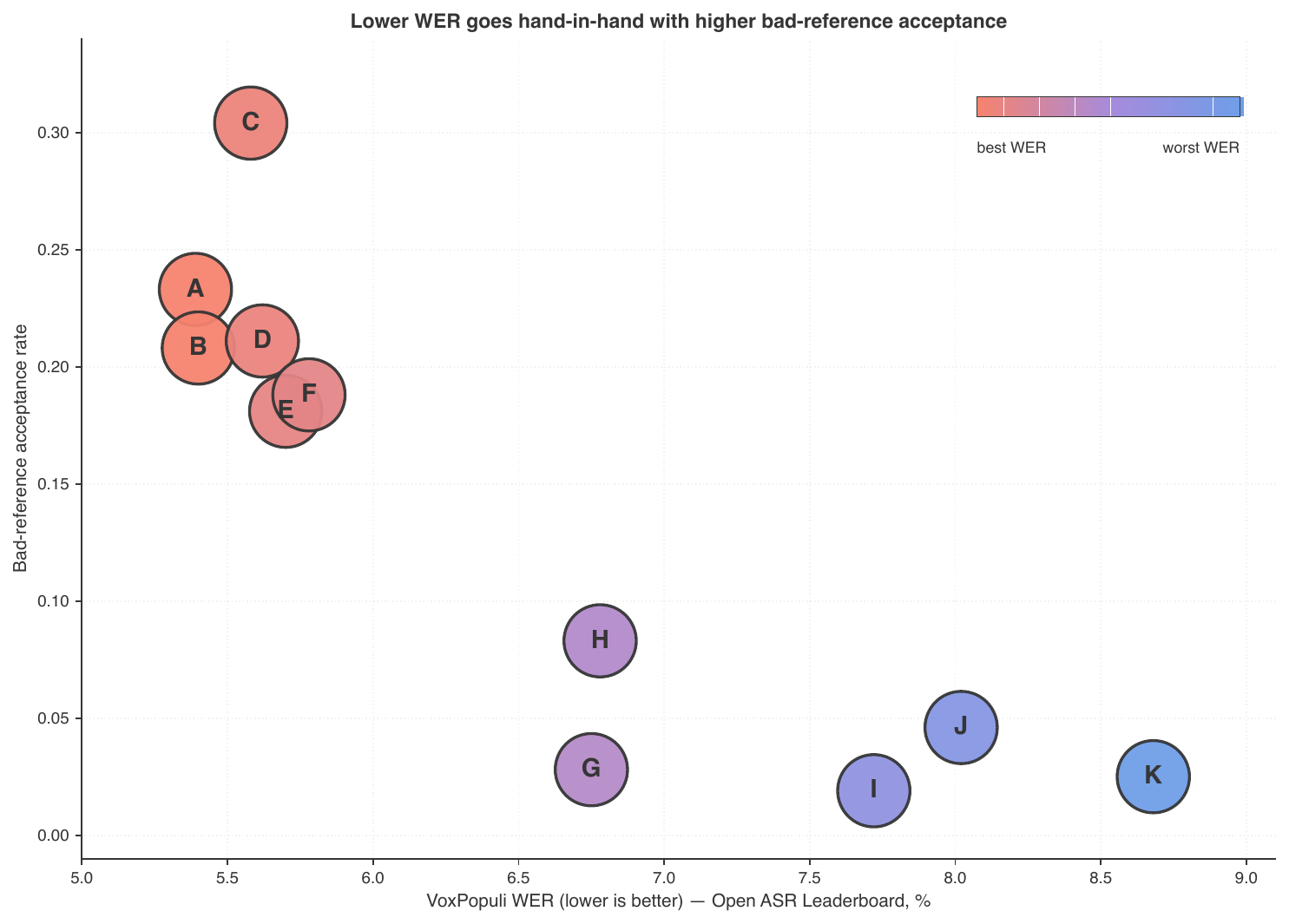}
    \caption{Cross-model audit on VoxPopuli-English. WER (\%) from the June 2026 Open ASR Leaderboard~\cite{srivastav2025openasrleaderboardreproducible}. Bad-reference acceptance: the rate at which each model returned reference transcripts verbatim on cases where the reference was identified as requiring a correction }
    \label{fig:benchmark-opt}
\end{figure}

We tested 11 models which currently are presented on the Hugginface Open ASR Leaderboard~\footnote{\href{https://huggingface.co/spaces/hf-audio/open_asr_leaderboard}{\texttt{huggingface.co/spaces/hf-audio/open\_asr\_leaderboard}}}~\cite{srivastav2025openasrleaderboardreproducible}. We show a select result from the VoxPopuli dataset in~\cref{fig:benchmark-opt}, future work will expand on these results.

The majority of models exhibit evidence of benchmark optimization on at least one diagnostic, including several of the top-performing systems on WER leaderboards. Importantly, the diagnostics implicate overlapping but non-identical sets of models, indicating that benchmark optimization manifests through multiple strategies rather than a single failure mode. Consensus disagreement and audio masking provide evidence of audio-level memorization: implicated models reproduce benchmark-specific transcripts despite contradictory or missing acoustic evidence, while the same content is transcribed correctly outside the benchmark setting. Orthographic switching captures a broader but less severe form of optimization, where models preferentially reproduce benchmark-specific annotation conventions for phonetically identical speech. Finally, the synthetic voice probe supports an audio-signature mechanism, with benchmark-specific behavior largely disappearing when the same content is rendered using a generic voice.

These preliminary results establish benchmark optimization as a measurable phenomenon across modern ASR systems and motivate the more detailed analyses that follow. In future work, we will extend this study and offer deeper analysis to better characterize the prevalence and mechanisms of benchmark optimization. More broadly, these findings reinforce the need for real world, withheld evaluation sets that remain inaccessible during model development, reducing opportunities for benchmark-specific optimization and providing a more faithful estimate of real world transcription performance. This motivation aligns with recent community efforts toward private-data evaluation, such as the Hugging Face Open ASR Leaderboard's
~\footnote{\href{https://huggingface.co/blog/open-asr-leaderboard-private-data}{\texttt{huggingface.co/blog/open-asr-leaderboard-private-data}}} introduction of hidden test sets. Our results suggest that such withheld evaluations are not merely desirable but necessary, as standard public benchmarks alone may substantially overestimate real world generalization.

\newpage

\section{Text-to-Speech Evaluation}
\label{sec:tts}

% \begin{figure}[H]
%     \centering
%     \includegraphics[width=\linewidth,trim=0 0 0 90pt,clip]{figures/Section_TTS_Opener_Hume.pdf}
%     \caption{Overview of the text-to-speech evaluation domain.}
%     \label{fig:tts-opener}
% \end{figure}

Text-to-speech (TTS) systems are increasingly expected to do more than synthesize intelligible, natural-sounding speech. Contemporary systems must preserve speaker identity, follow expressive and voice-design instructions, accurately render task-critical content, handle within-utterance language switching, remain stable during extended generation, and produce speech appropriate to the intended role and communicative setting. These capabilities can fail independently: a system may sound natural while drifting in speaker identity, produce expressive speech while misreading structured content, or perform well on short utterances while degrading over longer passages. \rwvoiceeq{} \textsc{Bench} therefore evaluates TTS as a multidimensional capability rather than reducing performance to a single naturalness or preference score. This section reviews prior TTS evaluation approaches, describes the \rwvoiceeq{} \textsc{Bench} TTS evaluation domain, presents dimension-level results, and discusses their implications for both TTS systems and automated evaluators.

\subsection{Experimental Setup}

We evaluated 31 TTS system configurations spanning commercial APIs, proprietary research systems, and open-source models. For each system, we used the provider's default voice, voice configuration, and generation settings. The evaluated systems were as follows; version details are provided in \cref{app:tts-models-used} and \cref{tab:tts-models-used}:

\begin{itemize}
    \item \textbf{OpenAI:} TTS-1, TTS-1-HD, and GPT-4o Mini TTS.
    \item \textbf{Google Gemini:} Gemini 2.5 Flash Preview TTS, Gemini 2.5 Pro Preview TTS, and Gemini 3.1 Flash TTS Preview.
    \item \textbf{ElevenLabs:} Eleven Multilingual v2 and Eleven v3, each evaluated through both standard TTS and voice-design interfaces.
    \item \textbf{Cartesia:} Sonic 3.5.
    \item \textbf{Deepgram:} Aura 2.
    \item \textbf{Inworld:} Inworld TTS 1, Inworld TTS 1 Max, Inworld TTS 2, and Inworld Voice Design.
    \item \textbf{Kokoro:} Hexgrad Kokoro-82M, evaluated through local, Modal, and Replicate deployments.
    \item \textbf{XTTS-v2:} evaluated through Modal and Replicate deployments.
    \item \textbf{Additional open-source and hosted models:} Microsoft VibeVoice-1.5B, Qwen3-TTS, Higgs Audio v2, Higgs Audio v3, Chatterbox, Dia, IndexTTS2, Parler-TTS, Fish Speech, and CSM-1B.
\end{itemize}

\textbf{Evaluation Tasks.} The TTS benchmark comprised 30 evaluations organized into seven evaluation dimensions: Acting / Role-Fit, Expressiveness, Voice Identity, Multilingual Code-Switching, Reliability, Long-Form Speaker Stability, and Acoustic Quality. Each category used a fixed item set and a predefined scoring protocol. Individual categories contained up to 120 prompts, although the number of applicable prompts varied according to the task design and the capabilities supported by each system.

\cref{tab:tts-evaluation-dimensions} summarizes these evaluation dimensions, task counts, rating volume, generation volume, and reported measures. The evaluations were designed to measure distinct constructs rather than interchangeable indicators of overall synthesis quality. For example, expressive tasks assessed whether a model realized a requested affective or pragmatic behavior, identity tasks assessed whether the perceived speaker remained stable under stress, and reliability tasks assessed whether task-critical lexical or symbolic content was rendered correctly. Results are therefore reported separately by dimension and are not combined into a single aggregate TTS score.

\begin{table}[H]
    \centering
    \scriptsize
    \setlength{\tabcolsep}{2pt}
    \renewcommand{\arraystretch}{1.15}
\caption{Summary of the \rwvoiceeq{} \textsc{Bench} TTS benchmark, including evaluation dimensions, constituent tasks, benchmark scale (prompts, generations, and human ratings), and reported evaluation measures.}
    \label{tab:tts-evaluation-dimensions}
    \begin{tabularx}{\textwidth}{@{}L{0.11\textwidth}Y r r r r Y@{}}
        \toprule
        \textbf{Dimension} & \textbf{Evaluation(s)} & \textbf{\#Prompts} & \textbf{\#Generations} & \textbf{\#Human Ratings} & \textbf{Avg Ratings/Gen} & \textbf{Reported measures \& scale} \\
        \midrule
        Acting / Role-Fit & Audiobook Narration, Sports Commentary, Stand-up Comedy, Humorous Delivery, Casual/Colloquial Reading, Vocal Bursts & 66 & 5,769 & 112,518 & 19.50 & Naturalness, prosody, role fit, character differentiation (1--5) \\

        Expressiveness & Inferred Emotion, Instructed Emotion, Pragmatic/Subtext Delivery, Sarcasm, Expressive Constraints, Mid-Utterance Emotional Transitions & 120 & 8,571 & 257,731 & 30.07 & Does the delivery realize the intended affective style, pragmatic meaning, expressive constraints, and emotion from prosody, rhythm, intensity, and voice quality? (1--5) \\
        
        Voice Identity & Emotional Stress, Neutral Baseline, Physical States, Prosodic Extremes, Register Shifts & 96 & 7,419 & 71,450 & 9.63 & Speaker consistency / same-speaker confidence (1--5) \\
        
        Language Stability & Code-Switching (DE/ES/FR/IT/JA), Heteronym Pronunciation & 70 & 6,150 & 140,960 & 22.92 & Native-accent perception and foreign-phrase comprehension (yes/no) \\
        
        Reliability & Pharmaceutical Terms, Alphanumeric Strings, Structured Data Reading/Spelling & 64 & 5,566 & 68,220 & 12.26 & Correct/clear read-back of hard content; exact-content accuracy (\%) \\
        
        Long-Form Speaker Stability & Long-Form Web Reading, Voice Identity -- Transitions, Voice Identity -- Long Form & 80 & 5,985 & 134,800 & 22.52 & Long-form naturalness and identity persistence through mid-clip shifts; audio-quality degradation opening to middle/closing (volume, noise, artifacts, pacing) (1--5, higher = less degraded) \\

        Acoustic Quality & Artifacts; Fidelity ratings collected within the expression evaluations & -- & -- & -- & -- & Naturalness, audio quality, absence of artifacts (1–5) \\

        \midrule
        \textbf{Overall} & \textbf{All TTS evaluation dimensions} & \textbf{496} & \textbf{39,460} & \textbf{785,679} & \textbf{19.91} & \textbf{Dimension-specific measures reported by dimension} \\
        \bottomrule
    \end{tabularx}
\end{table}

\textbf{Generation Protocol.} For each model–prompt pair, we generated three audio outputs through repeated calls to the corresponding system under the same task conditions.  Default provider settings were used unless an evaluation required a specific voice, reference input, language, speaking style, or voice-design instruction. Provider-specific prompt formatting was used only where necessary to express the same intended task through different system interfaces. Prompts and expected behaviors were fixed before model outputs were reviewed.

\textbf{Human Evaluation Protocol.} Human ratings were collected through Hume’s Study Runner API. For the primary English-language evaluations, participation was restricted to self-reported native English speakers located in the United States, United Kingdom, or Canada. Language stability evaluations used language-matched eligibility criteria corresponding to the target language.

Each audio clip was rated by at least three independent participants. Participants completed sessions containing no more than 10 clips, with one clip presented at a time. Sample order was randomized within each session, and participants were not shown model names, providers, deployment backends, or other system-identifying information.
Raters were shown only the contextual information required to apply the evaluation rubric. Depending on the task, this could include the source text, requested emotion or speaking style, intended role or genre, expected content, or a reference voice. Participants were instructed to use headphones and complete the study in an environment suitable for critical listening. 

Across all TTS studies, the annotation process yielded approximately 130,000 evaluated sample records and approximately 392,000 individual rater sessions. An evaluated sample record corresponds to one audio clip within one study after aggregating the ratings of three participants. An individual rater session corresponds to one participant evaluating one audio clip and completing all rating questions associated with that clip. The total number of criterion-level responses is therefore substantially larger than 392,000 because each rater session may contain multiple rating questions.

\textbf{Rating Rubrics and Scoring.} Rating questions and response scales were defined separately for each evaluation dimension. Most perceptual evaluations used five-point Likert scales, whereas correctness-oriented tasks used binary, categorical, or task-specific accuracy measures where appropriate. Acting and role-fit evaluations assessed dimensions such as naturalness, prosody, and genre appropriateness; expressiveness evaluations assessed affective or pragmatic fit, intensity, and delivery; voice-identity evaluations assessed perceived speaker consistency under expressive, physical, register, and duration-based stress; and acoustic-quality evaluations assessed artifacts, distortion, dropouts, instability, and other audible degradations. Complete model and evaluation metadata are provided in \cref{app:tts-models-used}.

For Likert-scale evaluations, ratings were first aggregated across the three participants assigned to each evaluated sample record. Repeated generations were retained as separate outputs and were subsequently summarized at the model–prompt and model–evaluation levels. Accuracy-based tasks were scored using predefined task-specific rules. Results are reported separately by construct and are not combined across incompatible scales into a single global TTS score.

\subsection{Results}

The TTS domain of \rwvoiceeq{} \textsc{Bench} evaluates seven evaluation dimensions: Expressiveness, Voice Identity, Multilingual Code-Switching, Reliability, Long-Form Speaker Stability, Acting / Role-Fit, and Acoustic Quality. Together, these evaluations assess whether a system can follow expressive instructions, preserve speaker identity, handle within-utterance language switching, render task-critical content reliably, remain stable during extended generation, maintain acoustic quality, and produce speech appropriate to a specified role or genre.

\begin{figure}[!t]
    \centering
    \includegraphics[width=0.7\linewidth,trim=0 1pt 0 5pt,clip]{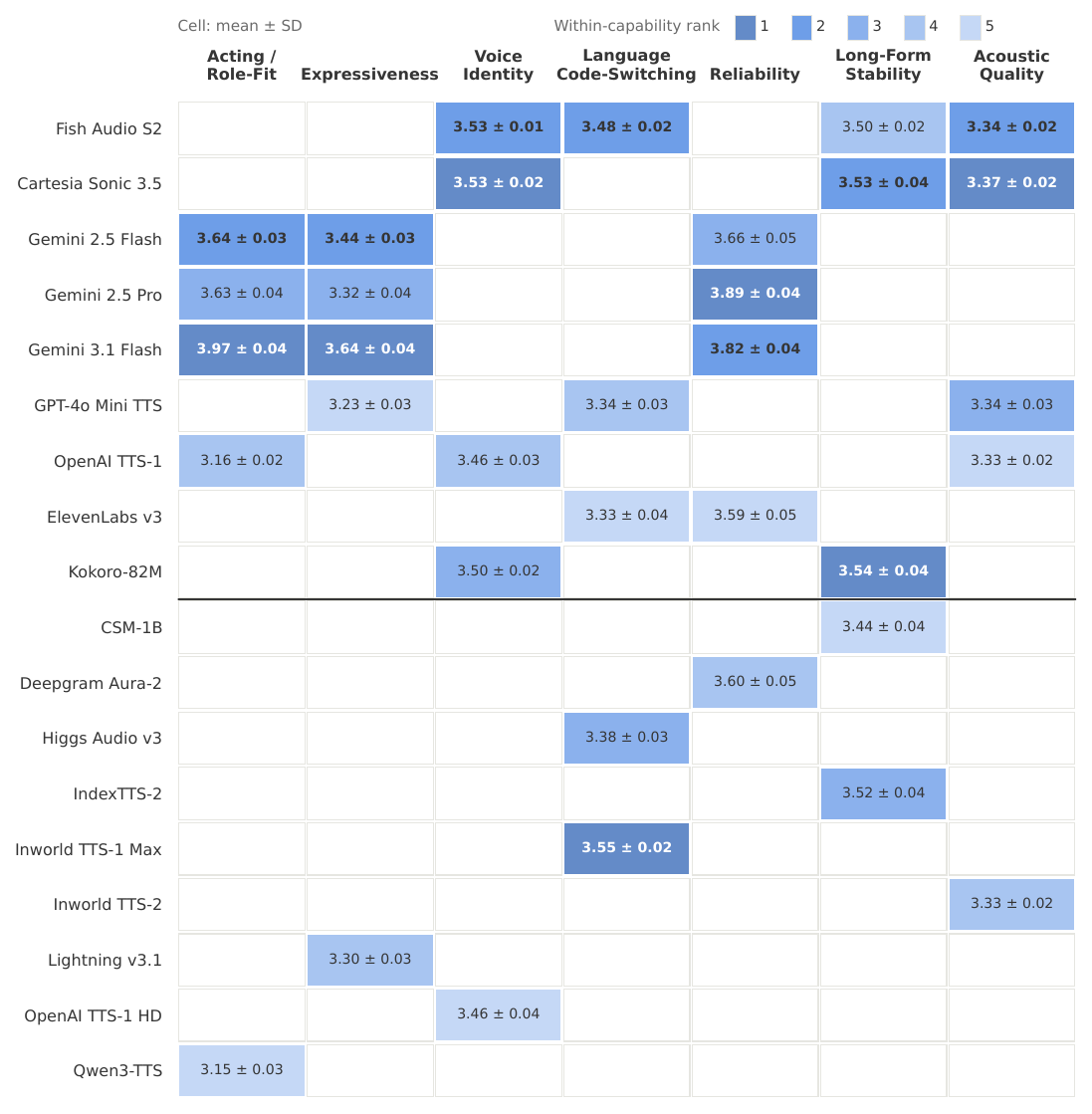}
    \caption{Top-five TTS system configurations by capability group. Cells report the capability-level mean $\pm$ standard deviation. Shading indicates within-capability rank, with darker cells denoting higher rank. Systems are ordered by the number of top-five appearances; the horizontal rule separates systems appearing in multiple top-five lists from those appearing in only one.}
    \label{fig:tts-top5-membership}
\end{figure}

\cref{fig:tts-top5-membership} presents the five highest-scoring system configurations within each evaluation dimension. Each cell reports the capability-level mean and standard deviation, while darker blue shading indicates a higher within-capability rank. Models are ordered by the number of top-five appearances. The resulting pattern shows substantial capability-specific variation: strong performance in one dimension does not consistently predict strong performance in another.

No system appears among the top five in all seven capability groups. Fish Audio S2 has the broadest coverage, appearing among the leading systems for Voice Identity, Language Stability, Long-Form Speaker Stability, and Acoustic Quality. The three Gemini configurations show broad but more concentrated performance: Gemini 3.1 Flash, Gemini 2.5 Flash, and Gemini 2.5 Pro each appear in the top five for Acting / Role-Fit, Expressiveness, and Reliability. Cartesia Sonic 3.5, GPT-4o Mini TTS, and OpenAI TTS-1 also appear in three capability groups, but with substantially different performance profiles.

The dimension leaders further illustrate these differences. Gemini 3.1 Flash leads Acting / Role-Fit ($3.97 \pm 0.04$) and Expressiveness ($3.64 \pm 0.04$), while Gemini 2.5 Pro leads Reliability ($3.89 \pm 0.04$). Voice Identity produces a different ordering: Fish Audio S2 and Cartesia Sonic 3.5 are tied at 3.53, followed by Kokoro-82M at 3.50 and OpenAI TTS-1 and TTS-1-HD at 3.46. Language Stability is led by Inworld TTS-1 Max ($3.55 \pm 0.02$), followed by Fish Audio S2 ($3.48 \pm 0.02$), indicating that performance on within-utterance language switching cannot be inferred from English-language expressiveness or acting performance.

Long-Form Speaker Stability produces another distinct ranking. After averaging the two long-form evaluation ranges, Kokoro-82M ranks first ($3.54 \pm 0.04$), followed closely by Cartesia Sonic 3.5 ($3.53 \pm 0.04$), IndexTTS2 ($3.52 \pm 0.04$), and Fish Audio S2 ($3.50 \pm 0.02$). Acoustic Quality is led by Cartesia Sonic 3.5 ($3.37 \pm 0.02$), with Fish Audio S2 and GPT-4o Mini TTS tied at 3.34. These results show that the systems leading expressive and instruction-following evaluations are not necessarily those that best preserve speaker identity, long-form stability, or acoustic quality.

Overall, the results support characterizing TTS systems through dimension-level performance profiles rather than a single aggregate score. A system that excels in expressive delivery may be less competitive in long-form speaker stability, while a system with strong identity preservation may perform less well on language stability or genre-specific acting. Collapsing these dimensions into a single mean would obscure materially different system profiles. Model selection is therefore use-case dependent: applications prioritizing continuity and content precision present a different selection problem from expressive narration, entertainment, or character-voice applications, for which acting, role fit, and expressive control may be more important.

\subsection{Discussion}

The TTS results show that high-quality generations must be more than natural. The strongest systems differ depending on whether the target is expressive delivery, speaker stability, reliability, pronunciation, or language stability.

Key takeaways:

\begin{itemize}
    \item Naturalness and expressiveness are distinct axes. A model can sound clean while failing to satisfy expressive, pragmatic, or role-specific instructions.
    \item Speaker identity is most vulnerable under expressive and physical stress. High arousal, whispering, loud projection, pitch shifts, register changes, and transitions between vocal states produce greater identity instability than complex linguistic content alone.
    \item Identity stability may coincide with under-realization of expressive instructions. Some systems preserve a consistent voice while producing only weak approximations of requested vocal actions such as laughter, screams, gasps, or abrupt emotional transitions. Identity preservation should therefore be considered together with expressive fidelity.
    \item Short-form quality does not guarantee long-form stability. Systems that perform strongly on isolated or short utterances may still exhibit speaker drift, pacing degradation, or reduced naturalness during extended generation.
    \item No single TTS model dominates all evaluation dimensions. Expressive and acting-oriented tasks favor one set of systems, while precision-focused tasks such as alphanumeric strings, structured data, and pharmaceutical terminology favor others.
\end{itemize}
\newpage

\section{Speech-to-Speech Evaluation}
\label{sec:sts}

Speech-to-speech systems must do more than generate a semantically appropriate response to a transcript. In spoken interaction, meaning is also conveyed through prosody, affect, hesitation, urgency, vocal effort, and acoustic context. A user may express fear while using neutral words, signal reluctance through tone despite verbally agreeing, or become difficult to understand because of noise or channel degradation. An effective voice agent must use these non-lexical cues when deciding how to respond, while also maintaining a natural and appropriate speaking style under conversational stress.

These requirements create evaluation challenges that are not captured by text-based dialogue benchmarks. A system may produce a strong response from the transcript alone while failing to benefit from the accompanying audio, or it may understand the user correctly while responding with an emotionally mismatched, unstable, or unnatural voice. \rwvoiceeq{} \textsc{Bench} therefore evaluates speech-to-speech systems across complementary dimensions of audio-conditioned understanding, emotion alignment, expressivity robustness, voice naturalness, and problem redirection. The goal is to separate response competence from the system’s ability to perceive and use information carried specifically by the spoken signal.

\subsection{Experimental Setup}

\textbf{Systems Evaluated.} We evaluated 15 speech-to-speech system configurations spanning native end-to-end speech models and modular voice-agent pipelines. Version details are provided in \cref{app:sts-models-used} and \cref{tab:sts-models-used}. The evaluated systems were:

\begin{itemize}
    \item Google Gemini: Gemini 2.5 Flash Native Audio Latest, Gemini 2.5 Flash Native Audio Preview 12-2025, and Gemini 3.1 Flash Live Preview.
    \item OpenAI: GPT Realtime 2, GPT Realtime Mini, and GPT Realtime 2025-08-28.
    \item Open-source end-to-end systems: Moshi, PersonaPlex, Kimi Audio 7B Instruct, and Qwen3-Omni 30B.
    \item Deepgram pipelines: Deepgram with GPT-4o and Deepgram with GPT-4o Mini.
    \item Inworld pipelines: Inworld with GPT-4o and Inworld with GPT-4o Mini.
    \item ElevenLabs: ElevenLabs Agent.
\end{itemize}

The Gemini, OpenAI, Moshi, PersonaPlex, Kimi Audio, and Qwen3-Omni configurations operate as native or integrated speech-in/speech-out systems. The Deepgram, Inworld, and ElevenLabs configurations use modular pipelines in which spoken input is transcribed or otherwise processed before an LLM generates a response that is rendered through a TTS component. We treat each complete deployed configuration as the unit of evaluation, since the domain measures the behavior of the full voice-agent system rather than the performance of its individual components.
%% VERIFY: confirm the cascade-index-to-LLM mapping below before compiling.
%% The 1/3 = GPT-4o, 2/4 = GPT-4o Mini assignment is inferred from the numbering
%% and must be checked against the scoring pipeline configs.
In the results and figures, the modular pipelines are referred to by cascade index: Deepgram Cascade-1 (GPT-4o) and Deepgram Cascade-2 (GPT-4o Mini) for the Deepgram pipelines, and Inworld Cascade-3 (GPT-4o) and Inworld Cascade-4 (GPT-4o Mini) for the Inworld pipelines.

\textbf{Evaluation Inventory.} The STS domain comprised five task families: ambiguous-tone interpretation, degraded-audio robustness, hostile-user behavior, urgent-user behavior, and emotion understanding. Together, these tasks evaluate whether voice agents use information carried by the spoken signal, remain effective under acoustic or behavioral stress, and preserve appropriate response content and vocal delivery.

The tasks differ in turn structure, input condition, and scoring protocol. Ambiguous-tone interpretation uses a short multi-turn interaction in which the user’s vocal expression conflicts with or qualifies the lexical content. Degraded-audio robustness tests recovery of task-critical information from telephony-like corrupted speech. The hostile-user and urgent-user tasks test behavioral stability and spoken delivery under difficult conversational conditions. Emotion understanding uses controlled input conditions to distinguish general response competence from sensitivity to vocal expression.

Emotion Alignment includes both the ambiguous-tone interpretation evaluation and a matched input-condition ablation. The primary evaluation measures whether an agent interprets vocal tone appropriately when it qualifies or conflicts with the user's words. As a complementary analysis, scenarios were presented under audio-only (AO) and transcript-only (TO) conditions with matched lexical content. The AO--TO comparison isolates whether access to prosody and other paralinguistic cues improves the agent's response beyond what can be inferred from the transcript.

\begin{table}[H]
    \centering
    \scriptsize
    \setlength{\tabcolsep}{2pt}
    \renewcommand{\arraystretch}{1.15}
\caption{Summary of the \rwvoiceeq{} \textsc{Bench} STS benchmark, including evaluation dimensions, constituent tasks, benchmark scale (prompts, generations, and human ratings), and reported evaluation measures.}
    \label{tab:sts-evaluation-dimensions}
    \begin{tabularx}{\textwidth}{@{}p{0.14\textwidth}X r r r r X@{}}
        \toprule
        \textbf{Dimension} & \textbf{Evaluation(s)} & \textbf{\#Prompts} & \textbf{\#Generations} & \textbf{\#Human Ratings} & \textbf{Avg Ratings/Gen} & \textbf{Reported measures \& scale} \\
        \midrule
        %% CHECK: the reported scale here is yes/no, but the results figure shows
        %% means of ~3.2--3.4 for this dimension. Either the scale annotation or
        %% the figure values need to be reconciled. The same applies to the
        %% Problem Redirection row (yes/no vs. means of ~3.5--3.8).
        Emotion Understanding & Ambiguous Tone -- Response & 18 & 288 & 4,071 & 14.14 & Does the agent respond appropriately to emotionally ambiguous audio (yes/no) \\
        
        % Emotion Alignment & Ambiguous Tone -- Interpretation, Audio-vs.-Transcript & 36 & 575 & 28,227 & 49.09 & Does the agent correctly interpret an ambiguous user tone; emotion-alignment rating on audio-only vs. transcript-only conditions (1--5) \\

        Emotion Alignment & Ambiguous-Tone Interpretation; Audio-vs.-Transcript Ablation & 36 & 575 & 28,227 & 49.09 & Does the agent correctly interpret an ambiguous user tone; matched AO--TO comparison (1--5) \\

        Expressivity Robustness & Degraded Line -- Task, Urgent Caller -- Reassurance, Hostile Caller -- Satisfaction & 14 & 568 & 9,031 & 15.90 & Task success under a poor phone line; reassurance/calming under urgency; leaving a hostile caller satisfied (yes/no, 1--5) \\
        
        Problem Redirection & Urgent Caller -- Decisiveness, Hostile Caller -- Composure, Hostile Caller -- Redirect & 10 & 400 & 3,555 & 8.89 & Decisive action on an urgent issue; maintaining composure and redirecting a hostile caller back to the problem (yes/no) \\
        
        Voice Naturalness & Degraded Line, Hostile Caller, Urgent Caller -- Voice Naturalness & 14 & 568 & 3,169 & 5.58 & Voice naturalness under degraded-line, hostile-load, and urgent-load conditions (1--5) \\

        \midrule
        \textbf{Overall (unique scenarios)} & \textbf{All STS evaluation dimensions} & \textbf{50} & \textbf{1,143} & \textbf{48,053} & \textbf{42.04} & \textbf{Dimension-specific measures reported by dimension} \\
        \bottomrule
    \end{tabularx}
    \begin{minipage}{\textwidth}
    \vspace{4pt}
    \footnotesize
    \emph{Note:} Constituent evaluations share underlying scenarios and generations across dimensions (e.g., Voice Naturalness is rated on the same generations as the Expressivity Robustness stress evaluations), so per-dimension prompt and generation counts do not sum to the Overall row, which counts unique scenarios and generations. Human-rating counts are disjoint and sum exactly to the Overall total.
    \end{minipage}
\end{table}

\textbf{Generation Protocol.} For tasks with repeated generations, systems were queried three times for each system–item pair under the same input and task conditions. For the ambiguous-tone and emotion-understanding evaluations, one response was generated per system–item condition. All generated responses were retained unless the request failed or the resulting output could not be evaluated.

User-side audio was fixed across systems within each task. The degraded, hostile-user, and urgent-user evaluations used prerecorded reference prompts, ensuring that differences between systems could not be attributed to variation in the input recording. Conversation scaffolding, task instructions, and condition-specific inputs were also held constant across systems except where interface-specific formatting was required.

Failed requests, empty responses, unusable audio, and unsupported system–task combinations were excluded from scoring and recorded as missing observations. No best-of-n response selection was performed.

\textbf{Evaluation Outputs.} STS responses were evaluated at two levels. First, task-specific content and behavioral criteria measured whether the agent understood the user and produced an appropriate response. Second, perceptual criteria measured the naturalness and suitability of the generated speech.

The exact rating dimensions varied by task. Ambiguous-tone and emotion-understanding tasks emphasized interpretation and emotional appropriateness; degraded-audio tasks emphasized information recovery; hostile-user tasks emphasized composure and redirection; and urgent-user tasks emphasized decisiveness and calming behavior. Voice naturalness was reported separately where applicable so that strong conversational behavior could be distinguished from high-quality spoken delivery.

\subsection{Results}

The speech-to-speech domain evaluates whether a voice agent can interpret spoken input and respond appropriately under realistic conversational pressure. Unlike text-only dialogue evaluation, STS evaluation must account for information carried by the acoustic signal, including affect, hesitation, urgency, vocal stance, and channel degradation. It must also assess the quality of the agent's spoken response. A transcript may preserve lexical content while omitting paralinguistic information that changes the appropriate response. \rwvoiceeq{} \textsc{Bench} therefore reports results separately for each STS evaluation dimension rather than collapsing them into a single aggregate score.

The STS domain comprises five evaluation dimensions:

\begin{itemize}
    \item \textbf{Emotion Understanding:} Tests whether the agent responds appropriately when the user’s vocal expression qualifies or conflicts with the lexical content of an emotionally ambiguous message. This dimension scores the appropriateness of the agent’s response behavior on such audio.
    \item \textbf{Emotion Alignment:} Tests whether the agent recognizes when vocal tone changes the meaning of an apparent confirmation. Across critical confirmation scenarios, the user may say “yes” while sounding hesitant, fearful, angry, confused, or reluctant. The agent must respond appropriately to the mismatch between lexical content and vocal stance. This dimension also contains the audio-only and transcript-only input ablations used to estimate whether access to the acoustic signal improves emotional interpretation beyond a transcript-only anchor.
    \item \textbf{Expressivity Robustness:} Tests whether the agent maintains task performance under conversational stress, including degraded audio, hostile callers, and urgent callers. This dimension measures whether the agent remains composed, decisive, and useful when the interaction is emotionally or acoustically difficult.
    \item \textbf{Voice Naturalness:} Tests whether the agent’s own voice remains natural in difficult conversational conditions. This dimension isolates the voice-naturalness sub-question from the same stress evaluations used for expressivity robustness, because vocal quality and task behavior can diverge.
    \item \textbf{Problem Redirection:} Tests whether the agent can remain composed with a hostile user and redirect the conversation back to the underlying problem. This dimension focuses on maintaining composure and steering the user toward resolution.
\end{itemize}

\begin{figure}[!t]
    \centering
    % ,trim=0 0 0 50pt,clip
    \includegraphics[width=\linewidth]{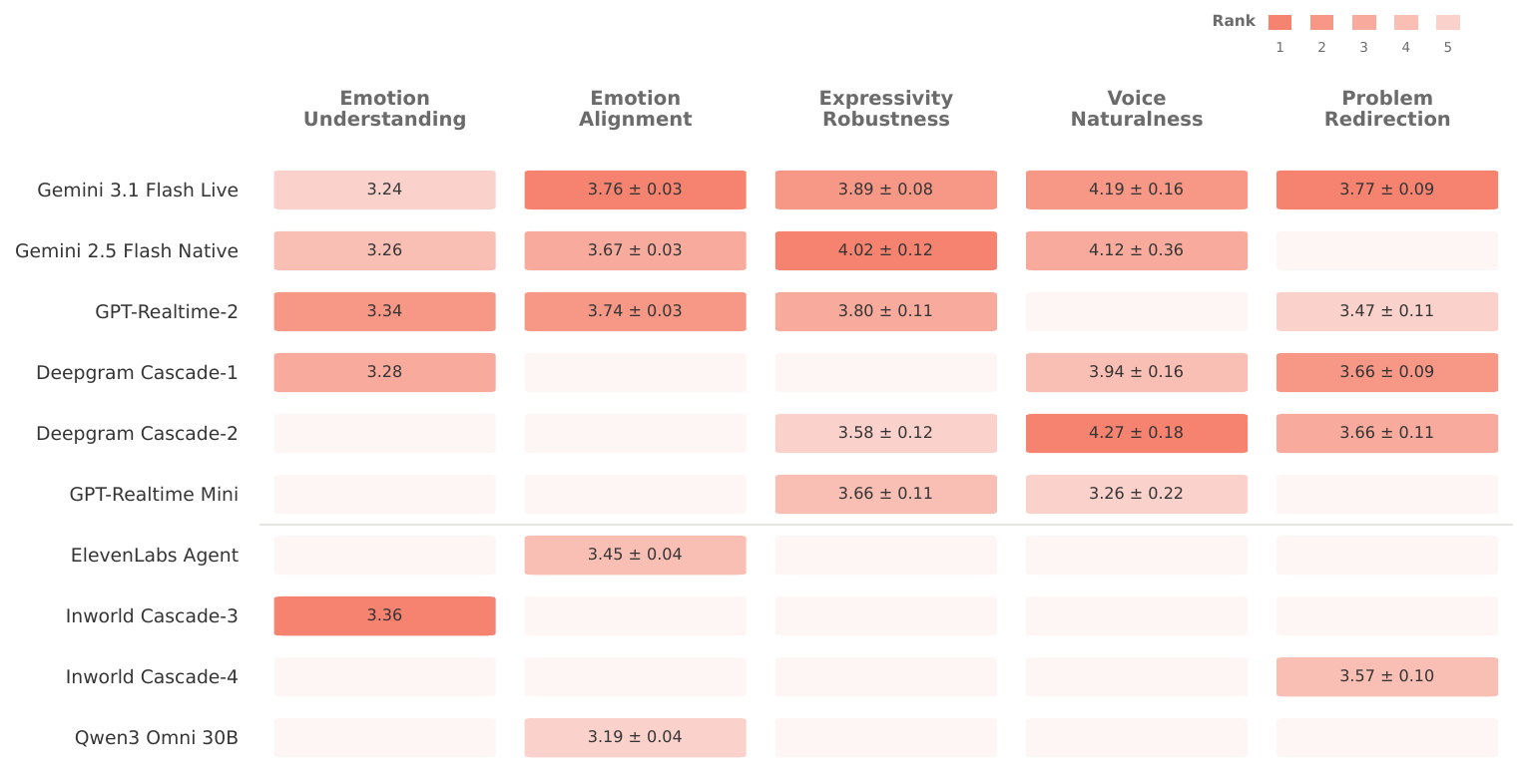}
    \caption{Top-performing speech-to-speech systems by evaluation dimension. Each column shows the five highest-scoring systems within an evaluation dimension, with cells reporting mean human rating ± standard deviation. Darker coral indicates a higher within-dimension rank (rank 1 darkest; rank 5 lightest), while blank cells indicate that the system did not place in the top five. Emotion Understanding maps to a single constituent evaluation, so no across-evaluation standard deviation is defined for that dimension. The horizontal line separates systems appearing in multiple top-five lists from those appearing in only one.}
    \label{fig:sts-top5-membership}
\end{figure}

\cref{fig:sts-top5-membership} shows the five highest-scoring systems in each dimension. Gemini 3.1 Flash Live has the strongest overall profile, leading Emotion Alignment and Problem Redirection and ranking second in both Expressivity Robustness and Voice Naturalness. GPT-Realtime-2 tracks it closely on the interaction-oriented dimensions but does not place in the top five for Voice Naturalness, an early indication that task behavior and vocal quality can diverge within a single system. The remaining dimensions are led by different systems: Gemini 2.5 Flash Native for Expressivity Robustness, Deepgram Cascade-2 for Voice Naturalness, and Inworld Cascade-3 for Emotion Understanding, while Inworld Cascade-4 enters the top five only for Problem Redirection, a pattern of dimension-specific specialization among the modular pipelines.

\paragraph{Audio-versus-transcript ablation.}
The audio-only (AO) and transcript-only (TO) conditions under Emotion Alignment test whether systems benefit from information carried by the acoustic signal. Across the 15 configurations, four systems scored higher under AO on the nine-scenario naturalistic set, one was unchanged, and 10 scored lower; on the 10-scenario scripted set, 10 scored higher and five scored lower. The three Gemini configurations remained close to zero on both sets, whereas GPT-Realtime-2 showed a consistent $+0.17$ AO--TO difference. Nine systems reversed the direction of the difference between the two sets. These descriptive results indicate that access to vocal expression is not uniformly beneficial and that its apparent value depends on both the system and the stimulus set.

Gemini 3.1 Flash Live has the broadest top-five coverage, appearing among the five highest-scoring configurations in all five dimensions. GPT-Realtime-2 and Gemini 2.5 Flash Native each appear in four dimensions, while Deepgram Cascade-1 and Deepgram Cascade-2 each appear in three. However, no system leads every dimension.

The dimension leaders further illustrate these differences. Emotion Understanding is led by Inworld Cascade-3 at 3.36, followed by GPT-Realtime-2 at 3.34 and Deepgram Cascade-1 at 3.28. Emotion Alignment is led by Gemini 3.1 Flash Live at 3.76 $\pm$ 0.03, narrowly followed by GPT-Realtime-2 at 3.74 $\pm$ 0.03. Expressivity Robustness produces a different ordering: Gemini 2.5 Flash Native ranks first at 4.02 $\pm$ 0.12, followed by Gemini 3.1 Flash Live at 3.89 $\pm$ 0.08 and GPT-Realtime-2 at 3.80 $\pm$ 0.11.

Voice Naturalness is led by Deepgram Cascade-2 at 4.27 $\pm$ 0.18, followed by Gemini 3.1 Flash Live at 4.19 $\pm$ 0.16 and Gemini 2.5 Flash Native at 4.12 $\pm$ 0.36. Problem Redirection is led by Gemini 3.1 Flash Live at 3.77 $\pm$ 0.09, with Deepgram Cascade-1 and Deepgram Cascade-2 tied at 3.66.

The variation in rankings indicates that task behavior and vocal quality are distinct aspects of STS performance. A system may respond appropriately under conversational stress while producing a comparatively unnatural voice, or it may generate natural-sounding speech without responding effectively to the user's underlying problem. Dimension scores are computed as unweighted means across their mapped constituent evaluations and are reported only for systems with complete coverage of the relevant dimension. The reported standard deviations describe variation across constituent evaluation-level scores; they are not estimates of rater-level uncertainty or confidence intervals. The Emotion Understanding score summarizes performance on its mapped evaluation but does not by itself establish whether audio provides an advantage over transcript input. That question must be assessed using the audio-only versus transcript-only contrasts included under Emotion Alignment.

\subsection{Discussion}

The STS results show that spoken-dialogue competence cannot be inferred from transcript-level dialogue quality. A system may produce strong responses from text while failing to use the appropriate tone, hesitation, urgency, or vocal affect.

Key takeaways:

\begin{itemize}
    %% NOTE: this first takeaway is only supported once the ablation paragraph
    %% above is filled in and uncommented. If the numbers cannot be included,
    %% soften this bullet (and the corresponding sentence in the abstract).
    \item Emotion-understanding ablations show that audio access is not consistently beneficial. Some systems remain largely transcript-driven, while others show measurable audio-over-text gain.
    \item Effective tone and emotion attunement requires both perception and response: recognizing the speaker’s vocal affect and adapting the system’s speech accordingly. These capabilities are complementary but distinct.
    \item Robustness under stress separates several abilities: handling degraded audio, remaining composed with hostile users, responding decisively to urgent users, and preserving natural spoken output.
    \item Voice naturalness and task behavior can diverge. A system may remain useful while sounding less natural, or sound natural while failing to redirect the interaction effectively.
\end{itemize}
\newpage

\section{Speech Understanding Evaluation}
\label{sec:su}

% \begin{figure}[H]
%     \centering
%     \includegraphics[width=\linewidth,trim=0 0 0 90pt,clip]{figures/Section_SU_Opener_Hume.pdf}
%     \caption{Overview of the text-to-speech evaluation domain.}
%     \label{fig:tts-opener}
% \end{figure}

Voice-agent evaluation increasingly relies on audio-language models as scalable substitutes for human listeners. These models are asked to judge whether generated speech is emotional, speaker-consistent, natural, or recognizably synthetic, often without first establishing that they can reliably perceive the corresponding acoustic attributes. This creates a fundamental validity problem: an automated judge may produce plausible and internally consistent scores while relying primarily on lexical content, dataset regularities, or broad audio cues rather than the specific perceptual construct under evaluation. For example, a model that cannot distinguish speakers should not be trusted to evaluate voice-identity preservation, and a model that cannot reliably compare emotional intensity should not be treated as an authoritative judge of expressive speech. The speech-understanding domain therefore evaluates the perceptual competence of candidate judges before using their outputs as evidence about voice-system quality.

Unlike conventional audio-understanding evaluation, the objective is not to construct a comprehensive leaderboard of general audio intelligence. Instead, \rwvoiceeq{} \textsc{Bench} asks a narrower methodological question: which models are sufficiently reliable for evaluating specific voice attributes, and when should an audio-language model be replaced by human ratings or a specialized acoustic model? The domain isolates four constructs that recur throughout voice-agent evaluation—emotion, relative affect, speaker identity, and synthetic-speech artifacts—and tests whether general-purpose audio-language models and task-specific systems can recover them directly from audio. These results provide the empirical basis for the benchmark’s hybrid evaluation strategy and prevent judge scalability from being mistaken for judge validity.

\subsection{Experimental Setup}

\paragraph{Systems Evaluated.}
We evaluated 18 speech-understanding systems as candidate audio judges, spanning
proprietary audio-language APIs, open-weight audio-language models, and dedicated
speaker-verification models. Version details are provided in \cref{app:su-models-used}
and \cref{tab:su-models-used}.

\begin{itemize}
  \item \textbf{Proprietary audio-language APIs (9).} Google Gemini 2.5 Pro,
  2.5 Flash, 2.5 Flash-Lite, 3.1 Pro Preview, 3.1 Flash-Lite, and 3.5 Flash;
  OpenAI gpt-audio, gpt-audio-mini, and gpt-audio-1.5. These accept audio
  directly and are queried through their respective vendor APIs.
  \item \textbf{Open-weight audio-language models (6).} Kimi-Audio-7B-Instruct,
  Phi-4-multimodal, Nemotron-3-Nano-Omni, Qwen3-Omni-30B, Voxtral-Small-24B, and
  Audio Flamingo~3. These were self-hosted from their public checkpoints on
  managed GPU infrastructure and served through an OpenAI-compatible interface.
  \item \textbf{Dedicated speaker-verification models (3).} TitaNet-Large,
  ECAPA-TDNN, and WavLM (base-plus-sv). These are not audio-language models; they
  produce speaker embeddings whose similarity is thresholded into a
  same/different decision, and they participate only in the speaker-matching
  evaluation.
\end{itemize}

Each system is evaluated on the task it is meant to perform. Audio-language models
are prompted with a fixed, task-specific instruction and return a structured
prediction. The judge instruction is part of each evaluation's identity, so any
change to the prompt re-runs the evaluation. Decoding is deterministic
(temperature~0), and one prediction is produced per (system, item). Systems
architecturally constrained to a single audio input per prompt (Audio Flamingo~3)
run the single-clip evaluations and the beep-merged \emph{combined} pair variants,
but not the separate-clip pair tasks. The speaker-verification models run only the
separate-clip speaker-matching task.

\paragraph{Evaluation Inventory.}
These four constructs are operationalized as six evaluations. The emotion and
speaker items are drawn from a prosodic-speech corpus that provides per-clip
emotion-score vectors, speaker identifiers, and transcripts; the
synthetic-detection items are real and TTS-generated recordings. Item counts, inputs, ground-truth sources, and metrics are summarized in \cref{tab:su-evals}, followed by descriptions of the individual evaluations.

\begin{table}[t]
  \centering
  \caption{Summary of the \rwvoiceeq{} \textsc{Bench} Speech-understanding evaluation, including evaluation dimensions, Input, ground truth, and reported evaluation measures.}
  \label{tab:su-evals}
  \small
  \begin{tabularx}{\linewidth}{@{}L{3.1cm} L{2.2cm} r Y Y@{}}
    \toprule
    Evaluation & Input & \#\,Items & Ground truth & Metric \\
    \midrule
    Emotion Identification & 1 clip & 150 & Corpus primary-emotion label (48-way vocabulary) & Exact-match accuracy \\
    Relative Emotion Comparison & 2 clips (A, B) & 200 & Clip with higher target-emotion score & Forced-choice accuracy \\
    Combined Emotion Comparison & 1 merged clip (A\,$\cdot$\,beep\,$\cdot$\,B) & 200 & (same as above) & Forced-choice accuracy \\
    Speaker Match & 2 clips (A, B) & 200 & Same- vs.\ cross-speaker pairing (100/100) & Forced-choice accuracy \\
    Combined Speaker Match & 1 merged clip (A\,$\cdot$\,beep\,$\cdot$\,B) & 200 & (same as above) & Forced-choice accuracy \\
    Synthetic-Speech Detection & 1 clip & 201 & Real vs.\ synthetic source (41/160) & Score gap; threshold accuracy \\
    \bottomrule
  \end{tabularx}
\end{table}

\begin{itemize}
  \item \textbf{Emotion Identification.} Whether a judge can identify the primary
  emotion expressed in a single clip from a large fixed emotion vocabulary; scored
  by exact-match accuracy against the corpus label.
  \item \textbf{Relative Emotion Comparison.} Whether a judge can decide which of
  two clips expresses a target emotion more strongly; this evaluates comparative
  affect perception rather than open-vocabulary labeling.
  \item \textbf{Combined Emotion Comparison.} The same pairwise task, but with both
  clips concatenated into a single beep-separated audio file, testing whether the
  judge can segment and compare two audio segments inside one input stream.
  \item \textbf{Speaker Match.} Whether two clips contain the same speaker or
  different speakers; includes both general audio-language judges and dedicated
  speaker-verification embedding models.
  \item \textbf{Combined Speaker Match.} The same same/different task with both
  clips presented as one combined recording; limited to audio-language judges,
  since the embedding models require separate audio inputs.
  \item \textbf{Synthetic-Speech Detection.} Whether a judge can distinguish real
  from synthetic speech; judges rate each clip on a 1--5 human-likeness scale.
\end{itemize}

\paragraph{Ground Truth.}
Consistent with the label-based scoring paradigm,
no human rater participates in the speech-understanding scoring loop. Model
predictions are compared directly against previously established references,
which are human-annotated samples. For
emotion identification, the reference is the corpus's primary-emotion label; for
the pairwise emotion comparison, the clip with the higher score on the target
emotion dimension; for speaker matching, the corpus same-/cross-speaker pairing;
and for synthetic-speech detection, the real/synthetic source of each clip. The
emotion and speaker corpora carry per-clip emotion-score vectors and speaker
identifiers used to derive these references deterministically; the
synthetic-detection set spans multiple clip durations ($\approx$30\,s, 1\,min,
3\,min, and longer conversational segments).

\paragraph{Scoring.}
Emotion identification is scored by exact-match accuracy against the corpus label
from a 48-category vocabulary with many semantically adjacent classes. The pairwise
emotion and speaker tasks are forced-choice and scored by accuracy against their
pairwise reference; the speaker-verification models decide via embedding
similarity. Synthetic-speech detection is rated on a 1--5 human-likeness scale: the
native metric is the mean human-likeness gap between real and synthetic clips, and
an illustrative threshold-derived accuracy is also reported by treating scores
above~3 as ``real'' and below~3 as ``synthetic'' (a rating of exactly 3 is counted
as incorrect).

\subsection{Results}

Speech understanding evaluates whether audio-language models and specialized audio models can extract perceptual, paralinguistic, speaker-level, and synthetic-speech information from audio. The input is audio, but the output is a decision: an emotion label, a relative affect comparison, a speaker-verification decision, or a real-versus-synthetic rating. Crucially, these tasks probe information carried by the sound signal itself, including prosody, timbre, speaker identity, and spectral artifacts, that a transcript does not preserve. This domain therefore evaluates whether models can understand how something was said, not merely what was said.

The speech-understanding results are organized around three evaluation dimensions:

\begin{itemize}
    \item \textbf{Emotion understanding:} Tests both categorical emotion identification from a single clip and relative emotion comparison between two clips, including a combined beep-separated variant. Together, these results separate open-vocabulary emotion labeling from comparative affect perception.
    \item \textbf{Speaker matching:} Tests whether systems can determine whether two clips contain the same speaker, using both separate-clip and combined-input variants. This result group compares general audio-language judges against dedicated speaker-verification embedding models.
    \item \textbf{Synthetic-speech detection:} Measures whether judges can distinguish real from synthetic speech using a 1--5 human-likeness scale. The native analysis reports real-versus-synthetic score gaps, with an illustrative threshold-derived accuracy also shown.
\end{itemize}

\begin{figure}[t]
    \centering
    \includegraphics[width=0.8\linewidth]{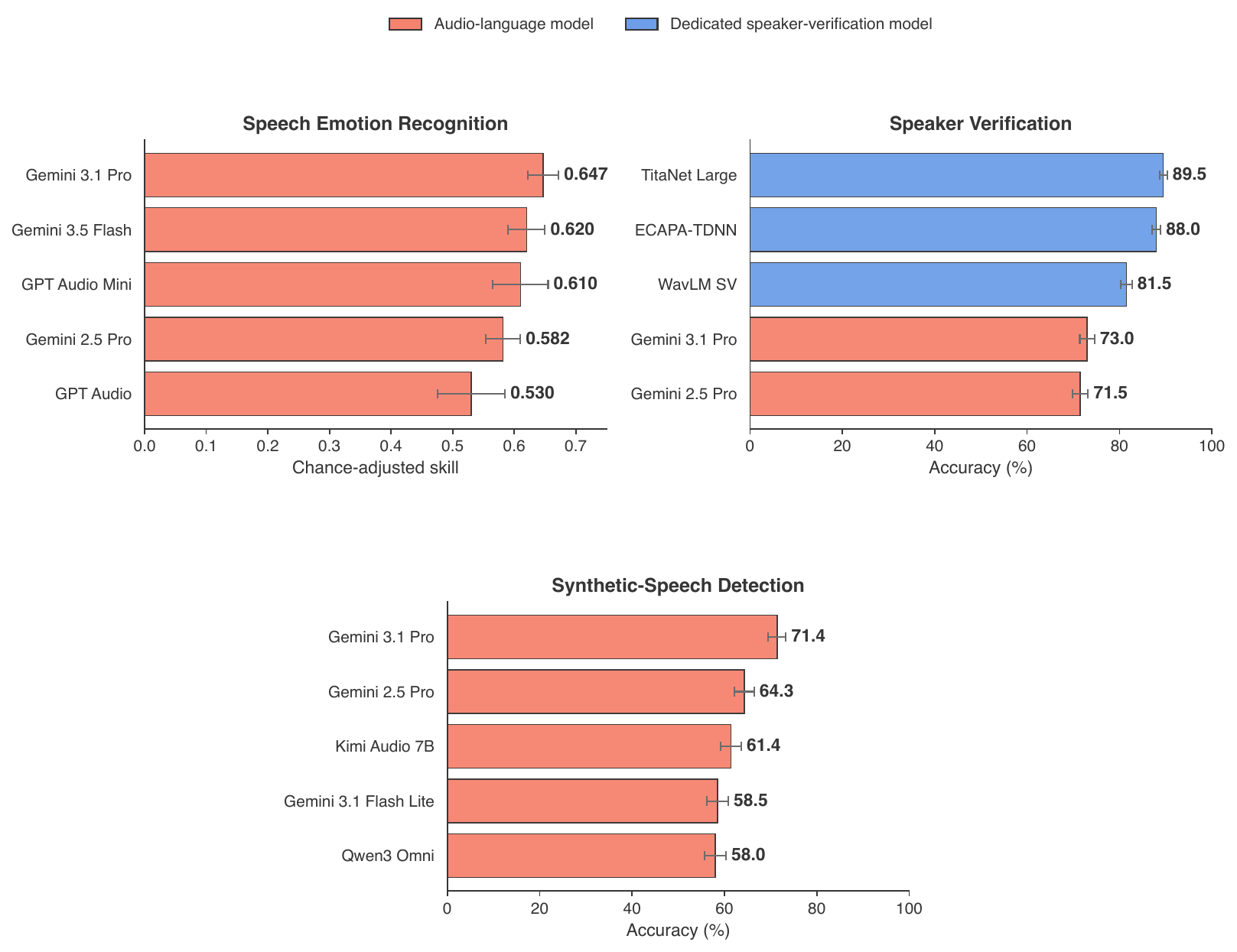}
  \caption{Top-performing speech-understanding systems across three evaluation dimensions. Panels show the five highest-scoring systems for speech-emotion recognition (chance-adjusted skill), speaker verification (accuracy), and synthetic-speech detection (threshold-derived accuracy). Colors distinguish general speech-language models from dedicated speaker-verification models.}
    \label{fig:su-top5-results}
\end{figure}

\Cref{fig:su-top5-results} reports the top five systems for each speech-understanding evaluation. Bars show accuracy for the forced-choice tasks. For synthetic-speech detection, bars show the illustrative implied accuracy obtained by thresholding ratings at $\geq$3 as “real”; the mean real-versus-synthetic score gap remains the more native metric for that evaluation. Dashed vertical lines indicate chance performance where applicable. Blue bars indicate systems in the overall speech-understanding top five, orange bars indicate task specialists that enter the top five for a specific evaluation but are not in the overall top five, and purple bars indicate dedicated speaker-verification models.
The results show that speech understanding is highly task-dependent. Absolute emotion identification is difficult for all judges: the best exact-match accuracy is only 22.7\%, despite 150 evaluated clips. This low ceiling is partly expected because the task requires selecting one label from a large emotion vocabulary with many semantically adjacent categories. Exact-match accuracy therefore likely underestimates partial perceptual competence, but it still shows that open-vocabulary emotion identification remains a hard task for current audio judges.

Relative emotion comparison is substantially easier. In the two-clip comparison task, Gemini 3.1 Pro Preview reaches 88.5\% accuracy, followed closely by Gemini 2.5 Pro and Gemini 3.5 Flash. The combined-input version produces a similar top ranking: the strongest Gemini models remain near 87–89\% accuracy even when both clips are presented as one beep-separated recording. This indicates that top-tier judges can make reliable relative affect judgments, even though they struggle with exact emotion naming. The contrast between emotion identification and relative emotion comparison is one of the clearest findings in the speech-understanding results.

The combined-input variants reveal a segmentation and comparison capability that is not captured by ordinary pairwise input. Strong Gemini models remain stable when clips are merged, but weaker models degrade substantially. For example, some open models drop to near chance or below chance in the combined emotion-comparison setting. This suggests that some judges can compare two separately supplied clips but cannot reliably parse “before” and “after” segments within a single continuous audio input. This distinction matters for practical evaluation pipelines, where judges may be asked to reason over concatenated or multi-segment audio.

Speaker matching shows the strongest specialized-versus-general contrast. Dedicated speaker-verification embedding models substantially outperform audio-language model judges: TitaNet Large SV reaches 89.5\%, ECAPA-TDNN reaches 88.0\%, and WavLM reaches 81.5\%, while the best general audio-language judge reaches 69.0\% on the separate-clip speaker-match task. This result indicates that general audio-language models should not be assumed to replace purpose-built speaker-verification systems when speaker identity is the target construct. In the combined speaker-match variant, where embedding models cannot be applied directly, Gemini 3.1 Pro Preview is the strongest audio-language judge at 73.0\%, followed by Gemini 2.5 Pro.

Synthetic-speech detection is also a distinct capability. Gemini 3.1 Pro Preview is the strongest discriminator, with the largest real-versus-synthetic score gap and the highest illustrative threshold accuracy. Kimi Audio and Gemini 2.5 Pro are the closest alternatives, but most judges assign high human-likeness scores to both real and synthetic clips. This suggests that strong performance on emotion perception or speaker comparison does not imply reliable synthetic-speech detection. Models that are useful as affect judges may still be easily fooled by synthetic audio.

Overall, speech understanding cannot be summarized by a single judge leaderboard. Gemini 3.1 Pro Preview is the most reliable all-round audio-language judge across the evaluated tasks, but specialized speaker-verification models dominate speaker matching, and many otherwise strong audio judges remain weak at synthetic-speech detection. The main conclusion is that audio-judge capability is construct-specific: relative emotion comparison, absolute emotion labeling, speaker verification, and synthetic-audio detection require different forms of perceptual competence.

\subsection{Discussion}

The speech-understanding results show that audio-language models are not universal audio judges. Emotion perception, speaker verification, and synthetic-speech detection require different capabilities which none of the models evaluated in this benchmark achieved in totality.

Key takeaways:
\begin{itemize}
    \item Absolute emotion identification remains weak, especially with a large and semantically overlapping emotion vocabulary.
    \item Models perform better when given a baseline or reference for comparing emotional states than when asked to assign emotion labels from a single utterance, suggesting that comparative affect perception is stronger than zero-shot categorical emotion recognition.
    \item Dedicated speaker-verification models outperform general audio-language judges on speaker matching, showing that specialized acoustic models remain necessary for speaker identity tasks.
    \item Synthetic speech detection and emotion understanding are distinct capabilities. Strong performance on emotion evaluation does not necessarily translate to reliable detection of synthetic speech; many judges still rate synthetic speech as being highly human.
\end{itemize}

\newpage

\section{Automatic Speech Recognition Evaluation}
\label{sec:asr}

% \begin{figure}[H]
%     \centering
%     \includegraphics[width=\linewidth,trim=0 0 0 135pt,clip]{figures/Section_ASR_Opener_Hume.pdf}
%     \caption{Overview of the automatic speech recognition evaluation domain.}
%     \label{fig:asr-opener}
% \end{figure}

Automatic speech recognition (ASR) has advanced rapidly in recent years, driven by increasingly capable foundation models and large-scale supervised and self-supervised training. As transcription systems become widely deployed in voice assistants, call centers, and conversational AI, they are expected to operate across a broad range of real world conditions, including diverse accents, noisy acoustic environments, and varied speaking styles and emotional expression. Consequently, robust evaluation has become as important as model development itself.

Despite this need, current leaderboard-style evaluations remain dominated by legacy public test sets that are both cleaner than production audio and increasingly vulnerable to contamination. Existing resources tend to isolate individual real world stressors such as accent  \cite{ardila2020commonvoice,wang2021voxpopuli}, noise and distance \cite{watanabe2020chime6,richey2018voices},emotion \cite{tuttosi2025bersting}, or spontaneity \cite{carletta2005ami, watanabe2020chime6}, rather than jointly modeling the conditions under which ASR is actually deployed: accented, emotionally expressive, conversational speech recorded under realistic acoustic conditions. Our benchmark addresses this gap directly by evaluating 40 models across four curated datasets that jointly capture accented speech, emotional expression, conversational interactions, and realistic background noise in a single unified evaluation, providing a more faithful assessment of ASR performance under real world conditions than existing clean or single-axis benchmarks~\footnote{The detailed view of the our ASR benchmark, is available at \href{https://huggingface.co/spaces/HumeAI/asr-leaderboard}{\texttt{huggingface.co/spaces/HumeAI/asr-leaderboard}}.}. 

\subsection{Experimental Setup}
We created four evaluation datasets targeting distinct real world conditions that are known to degrade ASR performance but are rarely represented in existing benchmarks: accented speech, emotionally expressive speech, speech with background noise or music, and naturalistic conversational speech. All datasets are English-only in this first release. Below we describe the data curation pipeline, the human annotation protocol used to establish ground-truth references, and the set of models evaluated.

\begin{itemize}
    \item \textbf{Accented speech.} Audio was sourced from publicly available online video content and automatically labeled into 23 accent categories using Gemini. A stratified subset of 230 clips (10 per accent, \textasciitilde1 hour) was selected for human annotation.
    \item \textbf{Emotional speech.} Audio was sourced from publicly available online video content and labeled using Hume AI's proprietary multilingual acoustic tagging model across 15 emotion categories. A stratified subset of 620 clips (\textasciitilde50 minutes) was selected for human annotation.
    \item \textbf{Background audio.} Audio was sourced from publicly available online video content and labeled using Hume's noise and music detection models. We selected a human-annotated subset of 460 clips (\textasciitilde40 minutes: 315 for noise, 145 for music).
    \item \textbf{Conversational speech.} For naturalistic dyadic conversation, we drew segments from the Hume-DaiKon corpus, a dataset of naturalistic dyadic interactions collected via a dual-channel conversational recording platform (Hume AI, 2026). We labeled 14,142 segments (15.4 hours) from more than 20 recording sessions.
\end{itemize}

Ground-truth references for the 4 different datasets were established through a two-stage human review process. For each audio, at least 3 independent raters reviewed and corrected model-generated transcriptions (and, for accents and emotions, the corresponding attribute label). Discrepancies between raters were resolved using a dedicated labeling/adjudication tool to produce a single consensus reference per clip.
Across all four datasets, this yields human-verified reference transcriptions (and, where applicable, attribute labels) against which all evaluated ASR systems are scored — distinguishing our benchmark from evaluations that rely solely on original source captions or single-pass automatic transcription as ground truth.

We evaluate 40 ASR systems spanning open-source and proprietary offerings from 16 organizations (\cref{tab:asr-models-used}). Open-source models include the NVIDIA Parakeet and Canary families, OpenAI Whisper (and its distilled and "lite" derivatives), IBM Granite Speech, Kyutai STT, Mistral Voxtral, Alibaba Qwen3-ASR, Cohere's transcription model, Microsoft Phi-4-multimodal, and Meta's MMS. Proprietary systems include ElevenLabs Scribe v2, Speechmatics Enhanced, AssemblyAI Universal-3 Pro, Deepgram Nova-3, Modulate Velma-2, Gladia Solaria, Smallest AI Pulse, Google Chirp 3, and Microsoft Azure Speech. This set spans a size range from lightweight streaming models (e.g., Parakeet TDT-CTC 110M) to large multi-billion-parameter systems, and includes both dedicated ASR architectures and general-purpose multimodal models applied to transcription.

We report Word Error Rate (WER) computed using the text normalization and scoring scripts from the Hugging Face Open ASR Leaderboard \cite{srivastav2025openasrleaderboardreproducible}, ensuring metric comparability with the most widely used public ASR leaderboard. For each of the four datasets we compute corpus-level WER; we additionally report an aggregate Average WER, defined as the unweighted arithmetic mean of the four per-dataset WERs — following the HF leaderboard convention of equal per-dataset weighting regardless of dataset size, so that no single dataset (e.g., the much larger conversational set) dominates the aggregate score. Inference speed is reported as RTFx (inverse real-time factor: audio duration divided by inference time), where higher values indicate faster transcription relative to audio length.

\subsection{Results}

\begin{figure}[!t]
    \centering
    % ,trim=0 0 0 50pt,clip
    \includegraphics[width=0.7\linewidth]{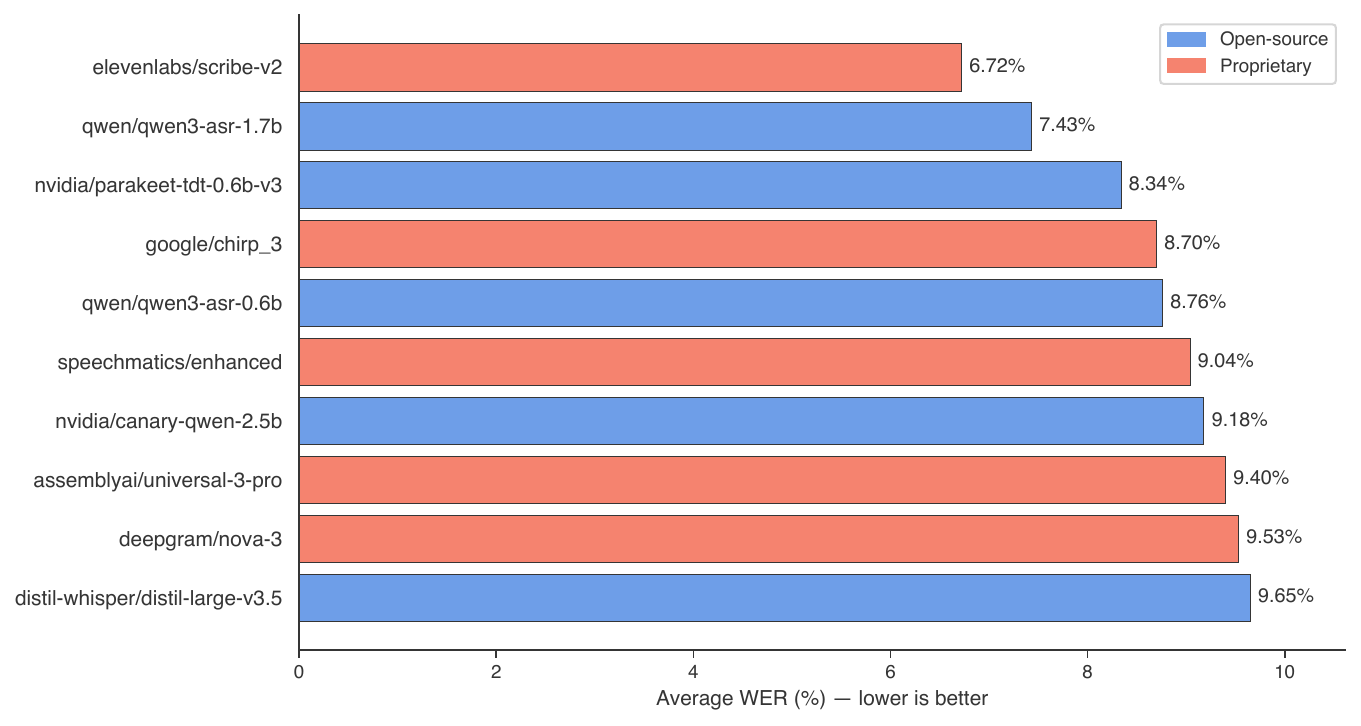}
    \caption{Performance of the top 10 ASR systems by average word error rate (WER) across four human-curated evaluation datasets. Colors distinguish proprietary and open-source models.}
    \label{fig:asr-top10}
\end{figure}

We evaluate 40 systems, including 14 open-source model families and 8 proprietary APIs, on four human-curated private datasets totaling roughly 17 hours of annotated audio.
% already in footnote above ^
% ~\footnote{https://huggingface.co/spaces/HumeAI/asr-leaderboard}.

% All results are browsable interactively at https://huggingface.co/spaces/HumeAI/hume-asr-leaderboard, with per-condition breakdowns and multiple accent grouping schemes.

The ASR robustness results summarize the main findings across evaluation conditions. ElevenLabs Scribe is the strongest overall system, with an average WER of 6.72\%, and leads on three of the four robustness tracks: accents, emotions, and noise/music. Qwen3-ASR-1.7B is the strongest open-source model, with an average WER of 7.43\%, and leads on conversational speech. Parakeet-TDT-0.6B-v3 provides the best speed–accuracy tradeoff, reaching 8.34\% average WER while running at 411× real time.

The accent results show that the native versus non-native gap is the most informative fairness axis. Across selected systems, the native-to-non-native WER gap ranges from +2.15 percentage points for ElevenLabs Scribe to +8.04 percentage points for AssemblyAI Universal-3 Pro. This gap is larger and more diagnostic than a standard-American versus other-regional-accents comparison, because the latter mixes easier native accents with harder non-native speech and hides the underlying fairness issue. Under Kachru's Three Circles taxonomy, the second-language (Outer Circle) group — Indian, South Asian, Nigerian, Filipino, and related varieties — is the hardest condition for many models. In this subgroup, the top ten systems fall between 6.58\% (Scribe) and 12.96\% (Modulate Velma-2) WER, while the same systems score between 4.38\% and 6.96\% on native English. Second-language English is therefore roughly 1.5–2× harder than native English even for otherwise strong systems, indicating substantial residual room for improvement.

Emotion and speaking style produce a different pattern. Positive high-arousal speech — amusement, excitement, laughter, and expressive rising intonation — is systematically harder to transcribe than negative high-arousal speech, across every model in the top ten. The positive-versus-negative gap runs from +3.25 to +5.14 percentage points for these systems: Scribe scores 7.81\% on positive high-arousal against 4.15\% on negative high-arousal (+3.66 pp), and Qwen3-ASR-1.7B — the strongest system on the emotion track after Scribe — scores 9.26\% on positive and 4.39\% on negative high-arousal (+4.87 pp). Negative low-arousal speech (sadness, grief, pain, boredom, anxiety) sits in between, ranging from 7.72\% to 11.83\% for the top ten. The consistency of the positive-versus-negative-high gap suggests that most training corpora underrepresent positive expressive speech: enunciation during joy and laughter deviates more from neutral read speech than anger and fear do, and models trained largely on neutral audio degrade accordingly.

Background audio also reveals an important split. Music-backed speech is close to solved for top systems, with music WER between 2.63\% (Scribe) and 3.96\% (Cohere Transcribe) for the top ten, while noise-backed speech remains much harder — 12.46\% to 19.51\% for the same systems. Across the evaluated 41 systems, noise WER is roughly four to five times higher than music WER (median 20.45\% on noise versus 4.84\% on music). A single averaged background-audio score therefore hides the real failure mode: music and babble noise are not equivalent robustness conditions. CrisperWhisper is the clearest outlier, with 57.20\% WER on noise compared with 7.33\% on music — a nearly 8× ratio, indicating that some architectures effectively factor out band-limited music but collapse under speech-like background noise.

Conversational speech reorders the ranking. Qwen3-ASR-1.7B leads this track with 6.45\% WER, ahead of Google Chirp-3 (7.05\%) and ElevenLabs Scribe (7.32\%) — a different top-3 than on the other three tracks, where Scribe is first. Two observations stand out. First, systems that couple acoustic modeling with an integrated language model — Qwen3-ASR, Canary-Qwen, and Chirp-3 — perform disproportionately better on spontaneous dialogue than on read speech, most likely because contextual language modeling absorbs the disfluencies, self-corrections, and colloquial constructions that characterize conversation. Second, the WER spread on this track is unusually wide: while the top ten systems fall between 6.45\% and 8.76\%, the median across all 41 evaluated systems is 9.64\% and the weakest reaches 31.54\%. Conversational speech is therefore the axis on which clean-benchmark performance is least predictive of real world behavior — the systems that win on read audio are frequently not the ones that will transcribe a natural conversation accurately.

Overall, the ASR results show that transcription robustness is condition-specific and that no single system dominates all four tracks. Scribe is the strongest on the more read-adjacent conditions (accents, emotions, background audio), Qwen3-ASR-1.7B is the strongest when spontaneous language modeling matters, and Parakeet-TDT-0.6B-v3 is the strongest when throughput matters. Strong performance on any one condition — including strong performance on standard clean-speech benchmarks — does not guarantee robustness to non-native accents, babble noise, positive expressive speech, or conversational delivery. ASR systems should therefore be evaluated through condition-level robustness profiles rather than a single WER score on clean benchmark corpora.

\subsection{Discussion}

Our ASR results argue against a single-number characterization of ASR quality. No system dominates every condition, and clean-benchmark WER does not reliably predict performance on non-native accents, noise-corrupted audio, expressive speech, or conversational dialogue.

Key takeaways:
\begin{itemize}
    \item \textbf{Accents:} Native-versus-non-native is a more informative fairness axis than regional accents. The 1.5--2$\times$ WER gap on second-language English likely reflects Inner-Circle bias in training data rather than an architecture problem.
    \item \textbf{Emotion:} Positive expressive speech (joy, laughter, excitement) is systematically harder than negative high-arousal speech, most likely because ASR training corpora underrepresent expressive styles that deviate from neutral read speech.
    \item \textbf{Background audio:} ``Noise'' is not one condition. Music is largely factorable; babble contains speech itself and cannot be removed by acoustic front-ends. The four-to-five-fold gap between the two means averaging them hides the failure mode that matters in production.
    \item \textbf{Conversational:} The leaderboard reorders on this track. Systems that couple acoustic modeling with an integrated language model (Qwen3-ASR, Canary-Qwen, Chirp-3) do best: contextual priors matter most for spontaneous, disfluent input.
    \item \textbf{Evaluation practice:} ASR systems should be reported through condition-level robustness profiles reflecting deployment context, not a single benchmark WER.
\end{itemize}

\newpage

\section{Discussion}

\rwvoiceeq{} \textsc{Bench} is the first benchmark to evaluate voice AI across speech generation, spoken interaction, audio understanding, and transcription robustness within a unified framework. Across these domains, we find that no single system consistently dominates every dimension. Instead, current frontier models exhibit increasingly specialized performance profiles: systems that excel at expressive speech generation may struggle with speaker identity, models that perform strongly on conversational reasoning may overlook vocal affect, and systems that achieve state-of-the-art transcription on clean benchmarks can degrade substantially under accented, expressive, noisy, or conversational speech.

Our findings suggest that performance is highly dimension-specific. In TTS, naturalness, expressiveness, identity preservation, and reliability diverge. In STS, access to audio does not guarantee that the agent uses vocal tone or affect. In speech understanding, relative audio judgments are much easier than fine-grained labeling, and specialized models still dominate speaker verification. In ASR, conventional clean-speech benchmarks do not capture the failures that arise in accented, emotional, noisy, and conversational speech.

Overall, \rwvoiceeq{} \textsc{Bench} advocates evaluating voice AI through domain-specific profiles rather than a single aggregate score. The benchmark exposes systems that are natural but not expressive, expressive but not identity-stable, transcript-competent but not audio-sensitive, or strong on clean transcription but brittle under real-world acoustic variation. As voice becomes a primary interface for AI, measuring these capabilities independently will become increasingly important for understanding how systems perform in deployment. Future work should expand sample sizes, incorporate paired statistical testing, refine partial-credit scoring for emotion labels, further disentangle acoustic perception from downstream response behavior in speech-to-speech systems, and continue investigating the role of speech-language models as scalable proxies for human evaluation.

\section{Conclusion}

We introduced Real World Voice EQ Bench, a multidimensional benchmark for evaluating modern voice AI systems across text-to-speech generation, speech-to-speech interaction, speech understanding, and ASR robustness. The benchmark is motivated by the observation that speech is not only text in audio form: it carries acoustic, expressive, speaker-level, and interactional information that affects how utterances are produced, interpreted, and acted on. Across the benchmark, we find that performance is highly domain-specific. In TTS, naturalness does not imply expressiveness, speaker-identity preservation, language switching, correctness, or long-form stability. In STS, access to audio does not guarantee that an agent uses vocal affect or tone, and some systems remain largely transcript-driven. In speech understanding, models perform unevenly across paralinguistic tasks, with relative emotion comparison easier than fine-grained emotion identification and dedicated speaker-verification models outperforming general audio-language models on speaker matching. In ASR, accented, emotional, noisy, and conversational speech expose production-relevant failures that are not captured by clean-speech benchmarks alone. These results support evaluating voice AI as a domain profile rather than a single aggregate score: a system may be expressive but not identity-stable, natural but not correct, transcript-competent but not audio-sensitive, or strong on clean transcription but brittle under real world acoustic variation. \rwvoiceeq{} \textsc{Bench} therefore provides a framework for measuring whether voice systems are not only intelligible or fluent, but genuinely sensitive to the acoustic, expressive, and interactional cues that shape spoken communication.

\section{Acknowledgements}
  We thank the Hume Engineering team for their support across infrastructure
  and design. In particular, we are grateful to Emily Beimfohr and Vince
  Picone (design logistics), Rob Hughes (talent coordination), Nick Palumbo
  (database infrastructure), and Nathan Leung (model infrastructure support), whose
  contributions made this work possible.

\bibliographystyle{unsrtnat}
\bibliography{sections/references}

@book{couperkuhlen2018interactional,
  author    = {Couper-Kuhlen, Elizabeth and Selting, Margret},
  title     = {Interactional Linguistics: Studying Language in Social Interaction},
  publisher = {Cambridge University Press},
  year      = {2018},
  doi       = {10.1017/9781139507318}
}

@book{barthweingarten2010prosody,
  editor    = {Barth-Weingarten, Dagmar and Reber, Elisabeth and Selting, Margret},
  title     = {Prosody in Interaction},
  series    = {Studies in Discourse and Grammar},
  volume    = {23},
  publisher = {John Benjamins},
  year      = {2010},
  doi       = {10.1075/sidag.23}
}

@book{gussenhoven2021handbook,
  editor    = {Gussenhoven, Carlos and Chen, Aoju},
  title     = {The Oxford Handbook of Language Prosody},
  publisher = {Oxford University Press},
  year      = {2021},
  doi       = {10.1093/oxfordhb/9780198832232.001.0001}
}

@book{laver1980voicequality,
  author    = {Laver, John},
  title     = {The Phonetic Description of Voice Quality},
  series    = {Cambridge Studies in Linguistics},
  volume    = {31},
  publisher = {Cambridge University Press},
  year      = {1980}
}

@book{scherer1979socialmarkers,
  editor    = {Scherer, Klaus R. and Giles, Howard},
  title     = {Social Markers in Speech},
  publisher = {Cambridge University Press},
  year      = {1979}
}

@article{sacks1974turntaking,
  author  = {Sacks, Harvey and Schegloff, Emanuel A. and Jefferson, Gail},
  title   = {A Simplest Systematics for the Organization of Turn-Taking for Conversation},
  journal = {Language},
  volume  = {50},
  number  = {4},
  pages   = {696--735},
  year    = {1974},
  url     = {https://www.jstor.org/stable/412243}
}

@article{chen2026voicebench,
  author  = {Chen, Yiming and Yue, Xianghu and Zhang, Chen and Gao, Xiaoxue and Tan, Robby T. and Li, Haizhou},
  title   = {{VoiceBench}: Benchmarking {LLM}-Based Voice Assistants},
  journal = {Transactions of the Association for Computational Linguistics},
  volume  = {14},
  pages   = {378--398},
  year    = {2026},
  doi     = {10.1162/tacl.a.628},
  url     = {https://aclanthology.org/2026.tacl-1.18/}
}

@article{liu2025vocalbench,
  author  = {Liu, Heyang and Wang, Yuhao and Cheng, Ziyang and Wu, Ronghua and Gu, Qunshan and Wang, Yanfeng and Wang, Yu},
  title   = {{VocalBench}: Benchmarking the Vocal Conversational Abilities for Speech Interaction Models},
  journal = {arXiv preprint arXiv:2505.15727},
  year    = {2025},
  url     = {https://arxiv.org/abs/2505.15727}
}

@article{shah2024speechrobustbench,
  author  = {Shah, Muhammad A. and Solans Noguero, David and Heikkil{\"a}, Mikko A. and Raj, Bhiksha and Kourtellis, Nicolas},
  title   = {Speech Robust Bench: A Robustness Benchmark for Speech Recognition},
  journal = {arXiv preprint arXiv:2403.07937},
  year    = {2024},
  url     = {https://arxiv.org/abs/2403.07937}
}

@article{liu2026speechparaling,
  author  = {Liu, Ruohan and Yin, Shukang and Wang, Tao and Zhang, Dong and Zhuang, Weiji and Ren, Shuhuai and He, Ran and Shan, Caifeng and Fu, Chaoyou},
  title   = {{SpeechParaling-Bench}: A Comprehensive Benchmark for Paralinguistic-Aware Speech Generation},
  journal = {arXiv preprint arXiv:2604.20842},
  year    = {2026},
  url     = {https://arxiv.org/abs/2604.20842}
}

@techreport{itu1996p800,
  author      = {{International Telecommunication Union}},
  title       = {{ITU-T Recommendation P.800}: Methods for Subjective Determination of Transmission Quality},
  institution = {International Telecommunication Union},
  number      = {P.800 (08/96)},
  year        = {1996},
  url         = {https://www.itu.int/rec/T-REC-P.800-199608-I/en}
}

@techreport{itu2015bs1534,
  author      = {{International Telecommunication Union}},
  title       = {{ITU-R Recommendation BS.1534-3}: Method for the Subjective Assessment of Intermediate Quality Level of Audio Systems},
  institution = {International Telecommunication Union},
  number      = {BS.1534-3},
  year        = {2015},
  url         = {https://www.itu.int/rec/R-REC-BS.1534-3-201510-I/en}
}

@inproceedings{black2005blizzard,
  author    = {Black, Alan W. and Tokuda, Keiichi},
  title     = {The Blizzard Challenge -- 2005: Evaluating Corpus-Based Speech Synthesis on Common Datasets},
  booktitle = {Proceedings of Interspeech 2005},
  pages     = {77--80},
  year      = {2005},
  doi       = {10.21437/Interspeech.2005-72}
}

@inproceedings{huang2022voicemos,
  author    = {Huang, Wen-Chin and Cooper, Erica and Tsao, Yu and Wang, Hsin-Min and Toda, Tomoki and Yamagishi, Junichi},
  title     = {The {VoiceMOS} Challenge 2022},
  booktitle = {Proceedings of Interspeech 2022},
  pages     = {4536--4540},
  year      = {2022},
  doi       = {10.21437/Interspeech.2022-970}
}

@article{huang2024voicemos,
  author  = {Huang, Wen-Chin and Fu, Szu-Wei and Cooper, Erica and Zezario, Ryandhimas E. and Toda, Tomoki and Wang, Hsin-Min and Yamagishi, Junichi and Tsao, Yu},
  title   = {The {VoiceMOS} Challenge 2024: Beyond Speech Quality Prediction},
  journal = {arXiv preprint arXiv:2409.07001},
  year    = {2024},
  url     = {https://arxiv.org/abs/2409.07001}
}

@article{huang2025instructttseval,
  author  = {Huang, Kexin and Tu, Qian and Fan, Liwei and Yang, Chenchen and Zhang, Dong and Li, Shimin and Fei, Zhaoye and Cheng, Qinyuan and Qiu, Xipeng},
  title   = {{InstructTTSEval}: Benchmarking Complex Natural-Language Instruction Following in Text-to-Speech Systems},
  journal = {arXiv preprint arXiv:2506.16381},
  year    = {2025},
  url     = {https://arxiv.org/abs/2506.16381}
}

@article{manku2025emergenttts,
  author  = {Manku, Ruskin Raj and Tang, Yuzhi and Shi, Xingjian and Li, Mu and Smola, Alex},
  title   = {{EmergentTTS-Eval}: Evaluating {TTS} Models on Complex Prosodic, Expressiveness, and Linguistic Challenges Using Model-as-a-Judge},
  journal = {arXiv preprint arXiv:2505.23009},
  year    = {2025},
  url     = {https://arxiv.org/abs/2505.23009}
}

@inproceedings{walker1997paradise,
  author    = {Walker, Marilyn A. and Litman, Diane J. and Kamm, Candace A. and Abella, Alicia},
  title     = {{PARADISE}: A Framework for Evaluating Spoken Dialogue Agents},
  booktitle = {Proceedings of the 35th Annual Meeting of the Association for Computational Linguistics and the 8th Conference of the European Chapter of the Association for Computational Linguistics},
  pages     = {271--280},
  year      = {1997},
  doi       = {10.3115/976909.979652}
}

@inproceedings{ao2024sdeval,
  author    = {Ao, Junyi and Wang, Yuancheng and Tian, Xiaohai and Chen, Dekun and Zhang, Jun and Lu, Lu and Wang, Yuxuan and Li, Haizhou and Wu, Zhizheng},
  title     = {{SD-Eval}: A Benchmark Dataset for Spoken Dialogue Understanding Beyond Words},
  booktitle = {Advances in Neural Information Processing Systems},
  volume    = {37},
  year      = {2024},
  doi       = {10.52202/079017-1813}
}

@inproceedings{yan2025urobench,
  author    = {Yan, Ruiqi and Li, Xiquan and Chen, Wenxi and Niu, Zhikang and Yang, Chen and Ma, Ziyang and Yu, Kai and Chen, Xie},
  title     = {{URO}-Bench: Towards Comprehensive Evaluation for End-to-End Spoken Dialogue Models},
  booktitle = {Findings of the Association for Computational Linguistics: EMNLP 2025},
  pages     = {17211--17242},
  year      = {2025},
  doi       = {10.18653/v1/2025.findings-emnlp.933},
  url       = {https://aclanthology.org/2025.findings-emnlp.933/}
}

@article{jiang2026s2sarena,
  author  = {Jiang, Feng and Lin, Zhiyu and Liu, Yiyang and Xue, Liumeng and Bu, Fan and Du, Yuhao and Chen, Xiangying and Wang, Benyou and Li, Haizhou},
  title   = {{S2S-Arena}: Evaluating Paralinguistic Instruction Following in Speech-to-Speech Models},
  journal = {arXiv preprint arXiv:2503.05085},
  year    = {2026},
  note    = {Version 2},
  url     = {https://arxiv.org/abs/2503.05085}
}

@inproceedings{yang2024airbench,
  author    = {Yang, Qian and Xu, Jin and Liu, Wenrui and Chu, Yunfei and Jiang, Ziyue and Zhou, Xiaohuan and Leng, Yichong and Lv, Yuanjun and Zhao, Zhou and Zhou, Chang and Zhou, Jingren},
  title     = {{AIR}-Bench: Benchmarking Large Audio-Language Models via Generative Comprehension},
  booktitle = {Proceedings of the 62nd Annual Meeting of the Association for Computational Linguistics},
  pages     = {1979--1998},
  year      = {2024},
  doi       = {10.18653/v1/2024.acl-long.109},
  url       = {https://aclanthology.org/2024.acl-long.109/}
}

@inproceedings{wang2025audiobench,
  author    = {Wang, Bin and Zou, Xunlong and Lin, Geyu and Sun, Shuo and Liu, Zhuohan and Zhang, Wenyu and Liu, Zhengyuan and Aw, AiTi and Chen, Nancy F.},
  title     = {{AudioBench}: A Universal Benchmark for Audio Large Language Models},
  booktitle = {Proceedings of the 2025 Conference of the Nations of the Americas Chapter of the Association for Computational Linguistics: Human Language Technologies},
  pages     = {4297--4316},
  year      = {2025},
  doi       = {10.18653/v1/2025.naacl-long.218},
  url       = {https://aclanthology.org/2025.naacl-long.218/}
}

@article{lee2025ahelm,
  author  = {Lee, Tony and Tu, Haoqin and Wong, Chi Heem and Wang, Zijun and Yang, Siwei and Mai, Yifan and Zhou, Yuyin and Xie, Cihang and Liang, Percy},
  title   = {{AHELM}: A Holistic Evaluation of Audio-Language Models},
  journal = {arXiv preprint arXiv:2508.21376},
  year    = {2025},
  url     = {https://arxiv.org/abs/2508.21376}
}

@inproceedings{ardila2020commonvoice,
  author    = {Ardila, Rosana and Branson, Megan and Davis, Kelly and Kohler, Michael and Meyer, Josh and Henretty, Michael and Morais, Reuben and Saunders, Lindsay and Tyers, Francis and Weber, Gregor},
  title     = {Common Voice: A Massively-Multilingual Speech Corpus},
  booktitle = {Proceedings of the Twelfth Language Resources and Evaluation Conference},
  pages     = {4218--4222},
  year      = {2020},
  url       = {https://aclanthology.org/2020.lrec-1.520/}
}

@inproceedings{conneau2023fleurs,
  author    = {Conneau, Alexis and Ma, Min and Khanuja, Simran and Zhang, Yu and Axelrod, Vera and Dalmia, Siddharth and Riesa, Jason and Rivera, Clara and Bapna, Ankur},
  title     = {{FLEURS}: Few-Shot Learning Evaluation of Universal Representations of Speech},
  booktitle = {2022 IEEE Spoken Language Technology Workshop (SLT)},
  pages     = {798--805},
  year      = {2023},
  doi       = {10.1109/SLT54892.2023.10023141}
}

@inproceedings{watanabe2020chime6,
  author    = {Watanabe, Shinji and Mandel, Michael and Barker, Jon and Vincent, Emmanuel and Arora, Ashish and Chang, Xuankai and Khudanpur, Sanjeev and Manohar, Vimal and Povey, Daniel and Raj, Desh and Snyder, David and Subramanian, Aswin Shanmugam and Trmal, Jan and Yair, Bar Ben and Boeddeker, Christoph and Ni, Zhaoheng and Fujita, Yusuke and Horiguchi, Shota and Kanda, Naoyuki and Yoshioka, Takuya and Ryant, Neville},
  title     = {{CHiME-6} Challenge: Tackling Multispeaker Speech Recognition for Unsegmented Recordings},
  booktitle = {Proceedings of the 6th International Workshop on Speech Processing in Everyday Environments},
  pages     = {1--7},
  year      = {2020},
  doi       = {10.21437/CHiME.2020-1}
}

@misc{srivastav2025openasrleaderboardreproducible,

  title={Open ASR Leaderboard: Towards Reproducible and Transparent Multilingual and Long-Form Speech Recognition Evaluation},

  author={Vaibhav Srivastav and Steven Zheng and Eric Bezzam and Eustache Le Bihan and Nithin Koluguri and Piotr Żelasko and Somshubra Majumdar and Adel Moumen and Sanchit Gandhi},

  year={2025},

  eprint={2510.06961},

  archivePrefix={arXiv},

  primaryClass={cs.CL},

  url={https://arxiv.org/abs/2510.06961}

}

@inproceedings{panayotov2015librispeech,
  title = {LibriSpeech: An ASR Corpus Based on Public Domain Audio Books},
  author = {Panayotov, Vassil and Chen, Guoguo and Povey, Daniel and Khudanpur, Sanjeev},
  booktitle = {2015 IEEE International Conference on Acoustics, Speech and Signal Processing (ICASSP)},
  pages = {5206--5210},
  year = {2015},
  organization = {IEEE}
}

@inproceedings{wang2021voxpopuli,
  title = {VoxPopuli: A Large-Scale Multilingual Speech Corpus for Representation Learning, Semi-Supervised Learning and Interpretation},
  author = {Wang, Changhan and Riviere, Morgane and Lee, Ann and Wu, Anne and Talnikar, Chaitanya and Haziza, Daniel and Williamson, Mary and Pino, Juan and Dupoux, Emmanuel},
  booktitle = {Proceedings of the 59th Annual Meeting of the Association for Computational Linguistics and the 11th International Joint Conference on Natural Language Processing (Volume 1: Long Papers)},
  pages = {993--1003},
  year = {2021}
}

@article{tuttosi2025bersting,
  title   = {{BERSting} at the Screams: A Benchmark for Distanced, Emotional and Shouted Speech Recognition},
  author  = {Tutt\"os\'i, Paige and Dhillon, Mantaj and Sang, Luna and Eastwood, Shane and Bhatia, Poorvi and Dinh, Quang Minh and Kapoor, Avni and Jin, Yewon and Lim, Angelica},
  journal = {Computer Speech \& Language},
  year    = {2025},
  note    = {arXiv:2505.00059}
}

@inproceedings{richey2018voices,
  title     = {{Voices Obscured in Complex Environmental Settings (VOiCES) Corpus}},
  author    = {Richey, Colleen and Barrios, Maria A. and Armstrong, Zeb and Bartels, Chris and Franco, Horacio and Graciarena, Martin and Lawson, Aaron and Nandwana, Mahesh Kumar and Stauffer, Allen and van Hout, Julien and Gamble, Paul and Hetherly, Jeffrey and Stephenson, Cory and Ni, Karl},
  booktitle = {Proc.\ Interspeech 2018},
  pages     = {1566--1570},
  year      = {2018},
  doi       = {10.21437/Interspeech.2018-1454}
}

@inproceedings{carletta2005ami,
  title     = {The {AMI} Meeting Corpus: A Pre-announcement},
  author    = {Carletta, Jean and Ashby, Simone and Bourban, Sebastien and Flynn, Mike and Guillemot, Ma{\"e}l and Hain, Thomas and Kadlec, Jaroslav and Karaiskos, Vasilis and Kraaij, Wessel and Kronenthal, Melissa and Lathoud, Guillaume and Lincoln, Mike and Lisowska, Agnes and McCowan, Iain and Post, Wilfried and Reidsma, Dennis and Wellner, Pierre},
  booktitle = {Machine Learning for Multimodal Interaction: Second International Workshop, MLMI 2005, Revised Selected Papers},
  series    = {Lecture Notes in Computer Science},
  volume    = {3869},
  pages     = {28--39},
  year      = {2006},
  publisher = {Springer},
  doi       = {10.1007/11677482_3}
}

\appendix
\clearpage
\begin{center}
  \vspace*{1.5em}
  {\Large\bfseries\color{HumeInk}Appendix}\par
  \vspace{0.45em}
  {\color{HumeRule}\rule{0.42\linewidth}{0.6pt}}\par
  \vspace{1.2em}
\end{center}
\addcontentsline{toc}{section}{Appendix}

\section{Text-to-Speech Models Evaluated}
\label{app:tts-models-used}

\begingroup
\scriptsize
\setlength{\tabcolsep}{3pt}
\renewcommand{\arraystretch}{1.15}
\begin{longtable}{@{}p{0.18\textwidth}p{0.27\textwidth}p{0.13\textwidth}p{0.22\textwidth}p{0.12\textwidth}@{}}
\caption{TTS models evaluated in \rwvoiceeq{}.}\label{tab:tts-models-used}\\
\toprule
\textbf{Model Name} & \textbf{API} & \textbf{Open Weight / Proprietary} & \textbf{Voice} & \textbf{Release date} \\
\midrule
\endfirsthead
\toprule
\textbf{Model Name} & \textbf{API} & \textbf{Open Weight / Proprietary} & \textbf{Voice} & \textbf{Release date} \\
\midrule
\endhead
\midrule
\multicolumn{5}{r}{\emph{Continued on next page}}\\
\endfoot
\bottomrule
\endlastfoot
OpenAI TTS-1 & tts-1 & Proprietary & alloy & 2023-11-06 \\
OpenAI TTS-1-HD & tts-1-hd & Proprietary & alloy & 2023-11-06 \\
OpenAI GPT-4o Mini TTS & gpt-4o-mini-tts & Proprietary & alloy & 2025-03-20 \\
Gemini 2.5 Flash Preview TTS & gemini-2.5-flash-preview-tts & Proprietary & Kore & May 2025 \\
Gemini 2.5 Pro Preview TTS & gemini-2.5-pro-preview-tts & Proprietary & Kore & May 2025 \\
Gemini 3.1 Flash TTS Preview & gemini-3.1-flash-tts-preview & Proprietary & Kore & 2026-04-15 \\
ElevenLabs Multilingual v2 & eleven\_multilingual\_v2 & Proprietary & rachel & 2023-08-22 \\
ElevenLabs v3 & eleven\_v3 & Proprietary & rachel & 2025-06-05 \\
Cartesia Sonic 3.5 & sonic-3.5 & Proprietary & katie & 2026-06-16 \\
Deepgram Aura 2 & aura-2-thalia-en (family: aura-2; language: en) & Proprietary & thalia & 2025-04-15 \\
Inworld TTS 1 & inworld-tts-1 & Proprietary & Ashley & 2025-10-17 \\
Inworld TTS 1 Max & inworld-tts-1-max & Proprietary & Ashley & 2025-10-17 \\
Inworld TTS 2 & inworld-tts-2 & Proprietary & Ashley & 2026-05-05 \\
Kokoro-82M & hexgrad/Kokoro-82M & Open Weight & af\_heart & 2024-12-25 \\
XTTS-v2 & xtts-v2 & Open Weight & Claribel Dervla & 2023-11-06 \\
Microsoft VibeVoice-1.5B & microsoft/VibeVoice-1.5B & Open Weight & p225 & 2025-08-25 \\
Qwen3-TTS & Qwen/Qwen3-TTS-12Hz-1.7B-CustomVoice & Open Weight & Ryan & 2026-01-22 \\
Higgs Audio v2 & eustlb/higgs-audio-v2-generation-3B-base & Open Weight & Server default & 2025-07-22 \\
Higgs Audio v3 & bosonai/higgs-audio-v3-tts-4b & Open Weight & Server default & June 2026 \\
Fish Speech & fishaudio/s2-pro & Open Weight & reference & 2026-03-09 \\
Chatterbox & resemble-ai/chatterbox & Open Weight & default & 2025-05-29 \\
Dia & zsxkib/dia & Open Weight & default & 2025-04-22 \\
IndexTTS2 & lucataco/indextts-2 & Open Weight & reference & 2025-09-08 \\
Parler-TTS & andreasjansson/parler-tts & Open Weight & ``A female speaker delivers a clear, expressive speech in a quiet, high-quality recording.'' & 2024-08-08 \\
Parler-TTS Large (Replicate) & parler-tts-large & Open Weight & ``A female speaker delivers a clear, expressive speech in a quiet, high-quality recording.'' & 2024-08-08 \\
CSM-1B & lucataco/csm-1b & Open Weight & 0 & 2025-03-13 \\
Fish Speech & jichengdu/fish-speech & Open Weight & reference & 2026-03-09 \\
Qwen3-TTS & qwen/qwen3-tts & Open Weight & Ryan & 2026-01-22 \\
TADA & HumeAI/tada-3b-ml & Open Weight & default & 2026-03-10 \\
Smallest.ai Lightning & lightning\_v3.1 & Proprietary & avery & 2026-03-27 \\
\end{longtable}
\endgroup

\newpage

\section{Speech-to-Speech Models Evaluated}
\label{app:sts-models-used}

\begingroup
\scriptsize
\setlength{\tabcolsep}{3pt}
\renewcommand{\arraystretch}{1.15}
\begin{longtable}{@{}p{0.20\textwidth}p{0.31\textwidth}p{0.14\textwidth}p{0.20\textwidth}p{0.09\textwidth}@{}}
\caption{Speech-to-speech models evaluated in \rwvoiceeq{}.}\label{tab:sts-models-used}\\
\toprule
\textbf{Model Name} & \textbf{STS Model API} & \textbf{Open Weight / Proprietary} & \textbf{Voice} & \textbf{Release date} \\
\midrule
\endfirsthead
\toprule
\textbf{Model Name} & \textbf{STS Model API} & \textbf{Open Weight / Proprietary} & \textbf{Voice} & \textbf{Release date} \\
\midrule
\endhead
\midrule
\multicolumn{5}{r}{\emph{Continued on next page}}\\
\endfoot
\bottomrule
\endlastfoot
OpenAI GPT-Realtime 2 & gpt-realtime-2 & Proprietary & alloy & 2026-05-07 \\
OpenAI GPT-Realtime & gpt-realtime-2025-08-28 & Proprietary & alloy & 2025-08-28 \\
OpenAI GPT-Realtime Mini & gpt-realtime-mini & Proprietary & alloy & Oct 2025 \\
Gemini 2.5 Flash Native Audio & gemini-2.5-flash-native-audio-latest & Proprietary & (native audio, no explicit voice) & May 2025 \\
Gemini 2.5 Flash Native Audio Preview & gemini-2.5-flash-native-audio-preview-12-2025 & Proprietary & (native audio, no explicit voice) & Dec 2025 \\
Gemini 3.1 Flash Live Preview & gemini-3.1-flash-live-preview & Proprietary & (native audio) & 2026-03-26 \\
Deepgram Voice Agent (LLM = GPT-4o) & STT: nova-3; LLM: gpt-4o; TTS: aura-2-thalia-en (Deepgram Voice Agent API) & Proprietary & thalia & 2025-06-16 \\
Deepgram Voice Agent (LLM = GPT-4o Mini) & STT: nova-3; LLM: gpt-4o-mini; TTS: aura-2-thalia-en & Proprietary & thalia & 2025-06-16 \\
Inworld Realtime (LLM = GPT-4o) & LLM: openai/gpt-4o (Inworld Router); TTS: inworld-tts-2 & Proprietary & Ashley & \\
Inworld Realtime (LLM = GPT-4o Mini) & LLM: openai/gpt-4o-mini; TTS: inworld-tts-2 & Proprietary & Ashley & \\
ElevenLabs Conversational\newline AI Agent & elevenlabs-agent & Proprietary & agent-\allowbreak configured; overridable via voice\_id & \\
Kyutai Moshi (Moshiko) & kyutai/moshi (moshiko variant) & Open Weight & moshiko (male, single voice) & 2024-09-17 \\
Moonshot Kimi-Audio 7B Instruct & moonshotai/Kimi-Audio-7B-Instruct & Open Weight & kimi (built-in single voice) & 2025-04-25 \\
NVIDIA PersonaPlex 7B v1 & nvidia/personaplex-7b-v1 & Open Weight & NATF0 default (bundled: NATF0-3, NATM0-3, VARF0-4, VARM0-4) & 2026-01-20 \\
Qwen3-Omni 30B-A3B Instruct & Qwen/Qwen3-Omni-30B-A3B-Instruct & Open Weight & Ethan default\newline (also: Chelsie, Aiden) & September 2025 \\
\end{longtable}
\endgroup

\newpage
\section{Speech Understanding Models Evaluated}
\label{app:su-models-used}

\begingroup
\scriptsize
\setlength{\tabcolsep}{3pt}
\renewcommand{\arraystretch}{1.15}
\begin{longtable}{@{}p{0.19\textwidth}p{0.40\textwidth}p{0.21\textwidth}p{0.14\textwidth}@{}}
\caption{Speech-understanding models evaluated in \rwvoiceeq{}.}\label{tab:su-models-used}\\
\toprule
\textbf{Model Name} & \textbf{API} & \textbf{Open / Proprietary} & \textbf{Release date} \\
\midrule
\endfirsthead
\toprule
\textbf{Model Name} & \textbf{API} & \textbf{Open / Proprietary} & \textbf{Release date} \\
\midrule
\endhead
\midrule
\multicolumn{4}{r}{\emph{Continued on next page}}\\
\endfoot
\bottomrule
\endlastfoot
 gemini-2.5-pro & Google Gemini API & Proprietary & 2025-06-17 (GA) \\
 gemini-2.5-flash & Google Gemini API & Proprietary & 2025-06-17 (GA) \\
 gemini-2.5-flash-lite & Google Gemini API & Proprietary & 2025-07-22 (stable; preview 06-17) \\
 gemini-3.1-pro-preview & Google Gemini API & Proprietary & 2026-02-19 (preview) \\
 gemini-3.1-flash-lite & Google Gemini API & Proprietary & 2026-05-07 (GA) \\
 gemini-3.5-flash & Google Gemini API & Proprietary & 2026-05-19 (GA) \\
 gpt-audio & OpenAI API & Proprietary & 2025-08-28 \\
 gpt-audio-mini & OpenAI API & Proprietary & 2026-01-19 \\
 gpt-audio-1.5 & OpenAI API & Proprietary & 2026-04-23 \\
 phi-4-multimodal & Self-hosted (HF: microsoft/Phi-4-multimodal-instruct) & Open weight (MIT) & 2025-02-26 \\
 kimi-audio-7b-instruct & Self-hosted (HF: moonshotai/Kimi-Audio-7B-Instruct) & Open weight & 2025-04-25 \\
 audio-flamingo-3 & Self-hosted (HF: nvidia/audio-flamingo-3-hf) & Open weight (NVIDIA, non-commercial) & 2025-07-09 \\
 nemotron-3-nano-omni & Self-hosted (HF: nvidia/Nemotron-3-Nano-Omni-30B-A3B) & Open weight (NVIDIA Open Model) & 2026-04-28 \\
 qwen3-omni & Self-hosted (HF: Qwen/Qwen3-Omni-30B-A3B-Instruct) & Open weight (Apache-2.0) & 2025-09-22 \\
 gemma-3n-e4b & Self-hosted (HF: google/gemma-3n-E4B-it) & Open weight (Gemma license) & 2025-06-03 \\
 voxtral-small-24b & Self-hosted (HF: mistralai/Voxtral-Small-24B-2507) & Open weight (Apache-2.0) & 2025-07-15 \\
 wavlm-sv & Self-hosted (HF: microsoft/wavlm-base-plus-sv) & Open weight (MIT) & 2021-10 (WavLM paper) \\
 ecapa-tdnn-sv & Self-hosted (HF: speechbrain/spkrec-ecapa-voxceleb) & Open weight (Apache-2.0) & 2020 arch / 2021 checkpoint \\
 titanet-large-sv & Self-hosted (HF: nvidia/speakerverification\_en\_\allowbreak titanet\_large) & Open weight (NeMo) & 2021-10 (TitaNet paper) \\
\end{longtable}
\endgroup

\newpage
\section{ASR Models Evaluated}
\label{app:asr-models-used}

\begingroup
\scriptsize
\setlength{\tabcolsep}{3pt}
\renewcommand{\arraystretch}{1.15}
\begin{longtable}{@{}p{0.18\textwidth}p{0.40\textwidth}p{0.13\textwidth}p{0.15\textwidth}@{}}
\caption{ASR models evaluated in \rwvoiceeq{}.}\label{tab:asr-models-used}\\
\toprule
\textbf{Group / Vendor} & \textbf{Model} & \textbf{Open / Proprietary} & \textbf{Date} \\
\midrule
\endfirsthead
\toprule
\textbf{Group / Vendor} & \textbf{Model} & \textbf{Open / Proprietary} & \textbf{Date} \\
\midrule
\endhead
\midrule
\multicolumn{4}{r}{\emph{Continued on next page}}\\
\endfoot
\bottomrule
\endlastfoot
nvidia & parakeet-ctc-1.1b & Open-source & 2023-12-28 \\
nvidia & parakeet-rnnt-0.6b & Open-source & 2023-12-28 \\
nvidia & parakeet-rnnt-1.1b & Open-source & 2023-12-27 \\
nvidia & parakeet-tdt-0.6b-v3 & Open-source & 2025-08-04 \\
nvidia & parakeet-tdt-1.1b & Open-source & 2024-01-25 \\
nvidia & parakeet-tdt-ctc-110m & Open-source & 2024-09-17 \\
nvidia & stt\_en\_conformer\_ctc\_large & Open-source & 2022-04-09 \\
nvidia & stt\_en\_fastconformer\_transducer\_large & Open-source & 2023-06-08 \\
nvidia & canary-180m-flash & Open-source & 2025-03-11 \\
nvidia & canary-1b & Open-source & 2024-02-07 \\
nvidia & canary-1b-flash & Open-source & 2025-03-07 \\
nvidia & canary-1b-v2 & Open-source & 2025-08-04 \\
nvidia & canary-qwen-2.5b & Open-source & 2025-06-26 \\
openAI & whisper-large-v3 & Open-source & 2023-11-07 \\
openAI & whisper-large-v3-turbo & Open-source & 2024-10-01 \\
distil-whisper & distil-large-v3.5 & Open-source & 2024-12-05 \\
ibm-granite & granite-4.0-1b-speech & Open-source & 2026-02-27 \\
ibm-granite & granite-speech-3.3-2b & Open-source & 2025-04-28 \\
ibm-granite & granite-speech-3.3-8b & Open-source & 2025-04-14 \\
kyutai & stt-2.6b-en & Open-source & 2025-06-06 \\
mistralai & Voxtral-Mini-3B-2507 & Open-source & 2025-07-01 \\
efficient-speech & lite-whisper-large-v3 & Open-source & 2025-02-26 \\
efficient-speech & lite-whisper-large-v3-acc & Open-source & 2025-02-26 \\
usefulsensors & moonshine-streaming-medium & Open-source & 2026-01-06 \\
qwen & qwen3-ASR-0.6B & Open-source & 2026-01-28 \\
qwen & qwen3-ASR-1.7B & Open-source & 2026-01-28 \\
cohereLabs & cohere-transcribe-03-2026 & Open-source & 2026-03-24 \\
nyrahealth & CrisperWhisper & Open-source & 2024-08-29 \\
zai-org & GLM-ASR-Nano-2512 & Open-source & 2025-12-09 \\
microsoft & Phi-4-multimodal-instruct & Open-source & 2025-02-24 \\
facebook & mms-1b-all & Open-source & 2023-05-27 \\
elevenlabs & scribe\_v2 & Proprietary & 2026-01-09 \\
speechmatics & enhanced & Proprietary & 2023-03-01 \\
assemblyai & universal-3-pro & Proprietary & 2026-02-03 \\
deepgram & nova-3 & Proprietary & 2025-02-12 \\
modulate & velma-2 & Proprietary & 2026-01-20 \\
gladia & solaria & Proprietary & 2025-04-02 \\
smallestAI & pulse & Proprietary & 2026-05-01 \\
google & chirp\_3 & Proprietary & 2025-03-01 \\
microsoft & azure-speech & Proprietary & 2025-10-15 \\
\end{longtable}
\endgroup

\FloatBarrier
\section{Demographics for Human Raters}
\label{app:human-demographics}
\begin{table}[H]
\centering
\caption{Demographic distribution from Human raters that contributed to the Benchmark via the Hume Study Runner API. Age statistics: $n=9{,}676$, mean $=38.7$, median $\approx37$ years.}
\label{tab:participant-demographics}
\begin{tabular}{llr|llr}
\toprule
\textbf{Trait} & \textbf{Category} & \textbf{\%} &
\textbf{Trait} & \textbf{Category} & \textbf{\%} \\
\midrule
\multirow{6}{*}{Age} 
& 18--24 & 12.3 &
\multirow{6}{*}{Ethnicity}
& White & 68.3 \\
& 25--34 & 31.7 &
& Black & 11.5 \\
& 35--44 & 26.0 &
& Asian & 9.7 \\
& 45--54 & 16.3 &
& Mixed & 6.0 \\
& 55--64 & 9.8 &
& Other & 2.9 \\
& 65+ & 3.9 &
& Prefer not to say & 1.0 \\
\midrule
\multirow{4}{*}{Sex}
& Female & 53.1 &
\multirow{4}{*}{Country}
& United States & 45.8 \\
& Male & 46.0 &
& United Kingdom & 36.1 \\
& Prefer not to say & 0.4 &
& Canada & 11.6 \\
& Revoked/Expired & 0.5 &
& Other & 6.5 \\
\bottomrule
\end{tabular}
\vspace{0.5em}
\footnotesize
\end{table}

\end{document}